\documentclass[aps,prd,twocolumn,superscriptaddress,nofootinbib,10pt]{revtex4-1}
\usepackage[paperwidth=21cm,paperheight=29.7cm,top=2.54cm,bottom=2.54cm,left=2cm,right=2cm]{geometry}
\usepackage[colorlinks,linkcolor=blue,anchorcolor=blue,citecolor=blue,urlcolor=blue]{hyperref}
\usepackage{graphicx,subfigure}
\usepackage[figuresright]{rotating}
\usepackage{amsmath,amsfonts,amssymb,bm}
\usepackage{array,enumitem,multirow}
\usepackage{acronym}
\usepackage{makecell}
\newcommand{\be}{\begin{equation}}
\newcommand{\ee}{\end{equation}}
\newcommand{\bea}{\begin{eqnarray}}
\newcommand{\eea}{\end{eqnarray}}
\newcommand{\nn}{\nonumber}

\newcommand{\td}{\tilde}

\newcommand{\tdh}{{\td{h}}}
\newcommand{\ukm}{{\rm~km}}
\newcommand{\mSun}{{\rm~M_\odot}}
\newcommand{\cO}{{\cal O}}
\newcommand{\TQC}{MOE Key Laboratory of TianQin Mission, TianQin Research Center for Gravitational Physics \&  School of Physics and Astronomy, Frontiers Science Center for TianQin, Gravitational Wave Research Center of CNSA, Sun Yat-sen University (Zhuhai Campus), Zhuhai 519082, China.}

\newacro{GR}{general relativity}
\newacro{GW}{gravitational wave}
\newacro{MG}{modified gravity theory}
\newacro{BH}{Black hole}
\newacro{PN}{post-Newtonion}
\newacro{ppE}{parameterized post-Einsteinian}
\newacro{GCB}{galactic ultra-compact binary}
\newacro{SBHB}{stellar-mass black hole binary}
\newacro{MBHB}{massive black hole binary}
\newacro{BHB}{black hole binary}
\newacro{IMBHB}{intermediate-mass black hole binary}
\newacro{EMRI}{extreme mass ratio inspiral}
\newacro{IMRI}{intermediate mass ratio inspiral}
\newacro{SGWB}{stochastic gravitational wave background}
\newacro{MECO}{minimal energy circular orbit}
\newacro{FAR}{false alarm rate}
\newacro{CE}{Cosmic Explorer}
\newacro{ET}{Einstein Telescope}
\newacro{LISA}{Laser Interferometer Space Antenna}
\newacro{EdGB}{Einstein-dilaton Gauss-Bonnet}
\newacro{dCS}{dynamic Chern-Simons}
\newacro{SNR}{signal-to-noise ratio}
\newacro{FIM}{Fisher Information Matrix}
\newacro{ISCO}{innermost stable circular orbit}
\newacro{NSBH}{neutron star-black hole binary}
\newacro{MCMC}{Markov Chain Monte Carlo}
\newacro{QNM}{quasi-nomral mode}

\begin{document}

\title{Constraining the EdGB theory with higher harmonics and merger-ringdown contribution using GWTC-3}

\author{Baoxiang Wang}
\author{Changfu Shi}
\email{Email: shichf6@mail.sysu.edu.cn (Corresponding author)}
\author{Jian-dong Zhang}
\author{Yi-Ming Hu}
\author{Jianwei Mei}
\affiliation{\TQC}

\date{\today}

\begin{abstract}
In this paper, we revisit the problem of using gravitational wave data to test the Einstein-dilation-Gauss-Bonnet theory, by using nine selected gravitational wave events from GWTC-3. Compared with existing work, we are taking into account the higher harmonics more properly and we also study the contribution of the merger-ringdown data. Using the inspiral data alone, we find that the best result is from GW200115, giving $\sqrt{|\alpha|} < 1.1$ km, which is about 17\% tighter than the previous best result. We also notice the possible existence of a simple unexpected relation among the constraints from different events. Several combinations of the selected events give $\sqrt{|\alpha|} \leq 1.0$ km. The result is further improved when the merger-ringdown data is also included, using two phenomenological schemes, giving $\sqrt{|\alpha|} < 0.87$ km for GW200115 in the best case scenario.
\end{abstract}

\maketitle

\section{Introduction}

The detection of \acp{GW} from compact binaries has opened the door to study the nature of gravity and dark compact objects in the genuinely strong field and dynamical regime. By using the currently available \ac{GW} data \cite{LIGOScientific:2016aoc,LIGOScientific:2018mvr,LIGOScientific:2020ibl,LIGOScientific:2021usb,LIGOScientific:2021djp}, a variety of tests have been performed \cite{LIGOScientific:2016lio,LIGOScientific:2018dkp,LIGOScientific:2019fpa,LIGOScientific:2020tif,LIGOScientific:2021sio,Perkins:2021mhb, Wang:2021jfc,Niu:2021nic,Wang:2021ctl,Kobakhidze:2016cqh,Yunes:2016jcc}. For example, there have been theory agnostic tests, such as residual test \cite{LIGOScientific:2016lio}, inspiral-merger-ringdown consistency test \cite{Ghosh:2016qgn}, and searches for possible deviations from the \ac{PN} waveform \cite{Arun:2006yw}, there have been tests on specific topics, such as no-hair theorem \cite{Isi:2019aib}, \ac{GW} polarization \cite{Isi:2017equ}, graviton mass \cite{LIGOScientific:2016lio}, the spin-induced quadrupole moment \cite{Krishnendu:2017shb}, the \ac{GW} dispersion relation \cite{Mirshekari:2011yq}, extra dimension \cite{Yunes:2016jcc}, time-varying gravitational constant G(t) \cite{Vijaykumar:2020nzc} and the gravitational Lorentz invariance \cite{Niu:2022yhr}, and there have also been tests targeting different \ac{MG} \cite{Yunes:2016jcc,Perkins:2021mhb,Wang:2021jfc,Niu:2021nic,Wang:2021ctl,Zhao:2019suc,Okounkova:2021xjv,Kobakhidze:2016cqh,Jenks:2020gbt,Zhu:2022uoq,Wu:2021ndf,Wang:2021gqm,Wang:2020cub,Haegel:2022ymk,Gong:2021jgg,Du:2020rlx}. No evidence against \ac{GR} has been found so far.

For the existing tests, the full inspiral-merger-ringdown data has been used in the residual test, the inspiral-merger-ringdown consistency test and the polarization test, while the test on no-hair theorem only uses the ringdown data, and tests on the \ac{PN} waveform and the spin-induced quadrupole moment and almost all theory/topic specific tests only use the inspiral data. What's more, previous works using the inspiral data have only considered corrections to the dominant 22 mode or have assumed a universal correction for all harmonics. For some tests, especially the theory-specific tests, it is natural to expect that both the higher harmonics and the merger-ringdown data can make a difference. But taking these factors into consideration has to come supported with adequate waveform modeling in the theories being tested. A good example here is the test of the \ac{EdGB} theory \cite{Kanti:1995vq,Torii:1996yi,Nojiri:2005vv,Yagi:2012gp}.

\ac{EdGB} is a quantum gravity-inspired modified gravity theory featuring a dilation coupled to the Gauss-Bonnet invariant, through a dimensionful coupling constant $\alpha$. Since the coupling is dimensionful, one generally expects the constraint to be of the order $\sqrt{|\alpha|} < \cO(L)\,$, where $L$ is the typical curvature radius of the system under consideration \cite{Berti:2015itd}. Indeed, electromagnetic observation of the orbital decay of black hole low-mass x-ray binary A0620-00 has produced a constraint, $\sqrt{|\alpha|} < 1.9\ukm$ \cite{Yagi:2012gp}, which is of the expected order.

Black hole binaries in \ac{EdGB} generically have dipolar \ac{GW} emission and so the theory is expected to get strong constraints with \acp{GW} \cite{Yagi:2011xp}. The first estimation of the constraints on \ac{EdGB} with real \ac{GW} source information has been done using the \ac{FIM} method \cite{Yunes:2016jcc}. The amplitude correction to the waveform was later included but the calculation was still \ac{FIM} based \cite{Tahura:2019dgr}. The full Bayesian analysis of \ac{EdGB}  was performed firstly with GWTC-1 data in \cite{Nair:2019iur} and then with GWTC-2 data in \cite{Perkins:2021mhb}. Both works have selected events with small mass ratios (defined to be the major mass over the minor mass). However, systems with larger mass ratios tend to place stronger constraints on $\alpha$ \cite{Shi:2022qno,Tahura:2018zuq,Yagi:2011xp}. For such systems, significant contributions from higher harmonics are expected. The contribution of higher harmonics has been considered in \cite{Wang:2021jfc,Lyu:2022gdr} with large mass ratio events such as GW190814, but a universal \ac{EdGB} correction to all harmonics has been assumed. Recently, a \ac{ppE} waveform model has been constructed in which the relations among the corrections to different harmonics have been carefully calculated \cite{Mezzasoma:2022pjb}.

Including the merger-ringdown data in theory-specific tests is challenging due to the difficulties with numerical relativity simulation \cite{Okounkova:2019dfo,Okounkova:2019zjf,Okounkova:2020rqw} and \ac{QNM} \cite{Cardoso:2019mqo,McManus:2019ulj,Baibhav:2023clw,Bao:2019kgt,Glampedakis:2017dvb,Blazquez-Salcedo:2016enn} calculation in the corresponding \acp{MG}. For \ac{EdGB}, a test with inspiral and ringdown data has been carried out based on \acp{QNM} found in a spherically symmetric background \cite{Carson:2020cqb,Carson:2020ter}. For more general situations, several schemes have recently been proposed \cite{Bonilla:2022dyt} to model phase corrections beyond the inspiral stage, based on the structure of IMRPhenomD waveforms \cite{Husa:2015iqa,Khan:2015jqa}.

In this paper, we revisit the problem of testing \ac{EdGB} with real \ac{GW} data. We refine existing work in the following ways:
\begin{itemize}
\item We apply the work of \cite{Mezzasoma:2022pjb} to \ac{EdGB} and obtain a waveform model containing the correct relation among the corrections to different harmonics. To apply the result of \cite{Mezzasoma:2022pjb}, we need to use waveform models that assume nonprecessing binaries, for this we will use the IMRPhenomXHM waveform \cite{Garcia-Quiros:2020qpx}.
\item We apply the method of \cite{Bonilla:2022dyt} to the IMRPhenomXHM waveform so that the merger-ringdown data can also be used.
\item We have included a couple of new events from GWTC-3 that has never been used to test \ac{EdGB} before.
\end{itemize}
We find that: (i) the appropriate inclusion of higher harmonic modes can make appreciable difference in the result. For example, there can be about 50\% improvements on the results from GW190707, GW190720 and GW190728 over previous ones that do not include higher harmonics \cite{Perkins:2021mhb}, and there can be nearly $30\%$ improvement from GW190421 compared to the previous result that assumes a universal correction for all higher harmonics \cite{Wang:2021jfc}; (ii) All constraints from the selected \ac{GW} events seem to be closely distributed by the line described by (\ref{eq.fit}) below; (iii) the effect of merge-ringdown data is relatively less significant, but can still give about 20\% improvement for GW200115, compared to that without using the merger-ringdown data; (iv) Our best constraint comes from GW200115, which with higher harmonic modes alone gives $\sqrt{|\alpha|}<1.1\ukm$, and if the merge-ringdown data is also taken into account, it gives $\sqrt{|\alpha|}<0.87\ukm$. As a comparison, the current best constraint on \ac{EdGB} comes from \cite{Lyu:2022gdr}, which also uses GW200115, giving $\sqrt{|\alpha|}<1.33\ukm$.

This paper is organized as follows. We introduce how the higher harmonics and the merger-ringdown contribution are included into the waveform model in Section \ref{sec:method}, select the \ac{GW} events to be used and make necessary explanations of the statistical method in Section \ref{sec:model}, present the main results in Section \ref{sec:re}, and then make a short concluding remark in Section \ref{sec:con}. 

We use the convention G=c=1 throughout the paper.


\section{Waveform models}\label{sec:method}

In this section, we explain the construction of the waveform models used in this paper.

\subsection{The ppE waveform model with higher harmonic contributions}\label{sub:hm}

The evolution of compact binaries is divided into three stages: inspiral, merger and ringdown. During the inspiral stage, the binary components are widely separated, their velocities are relatively small, and the \ac{PN} approximation \cite{Blanchet:2013haa} can be used to obtain the waveforms for low mass ratio systems. The \ac{ppE} waveform model has been proposed to capture the common features of how the \ac{PN} waveforms in many \acp{MG} deviate from those in \ac{GR} \cite{Yunes:2009bv}. Keeping only the leading order correction, the inspiral waveform for an \ac{MG} can be written as:
\be\tdh_{\rm ppE}(f)=\tdh_{\rm GR}(f)(1+\alpha u^a)e^{i\beta u^b}\,,\label{eq:waveform}\ee
where $\alpha$ and $\beta$ are the \ac{ppE} parameters,  $b=2{\rm~ PN}-5$ and $a=b+5$ are the \ac{ppE} order parameters, and the $\rm~ PN$ in this equation symbolizes the PN order,  $u=(\pi\mathcal{M}f)^{1/3}$ is a characteristic velocity, $\mathcal{M}=\eta^{3/5}M$ is the chirp mass,  $M=m_1+m_2$ is the total mass, $\eta=m_1m_2/(m_1+m_2)^2$ is the symmetrical mass ratio, $m_1$ and $m_2$ are the major and minor masses, respectively, and $h_{\rm GR}$ is the corresponding \ac{GR} waveform, which will be produced using IMRPhenomXHM in this paper.

To take into account the contribution of higher harmonics, the construction of \cite{Mezzasoma:2022pjb} neglects the contribution of the amplitude correction, leading to
\be\tdh_{\rm ppE}(f)=\tdh_{\rm GR}(f)e^{i\beta u^b}\,.\label{eq:phase}\ee
The error introduced in this process is expected to be less than a few percent \cite{Tahura:2019dgr}. Early works \cite{Nair:2019iur,Perkins:2021mhb} of using GWTC-1 and GWTC-2 data to test \ac{EdGB} have only considered the 22 mode of $\tdh_{\rm GR}$ in (\ref{eq:phase}), while the authors of \cite{Wang:2021jfc,Lyu:2022gdr} have considered the contribution of higher harmonics but have assumed a universal \ac{EdGB} correction to all harmonics, i.e.,
\be\tdh_{\rm ppE}(f)=\Big[\sum_{\ell m}\tdh_{\ell m}^{\rm GR}(f)\Big]e^{i\beta u^b}\,.\label{eq:phase2}\ee
However, there is no reason to believe that the \ac{ppE} corrections to all the harmonics are the same, and one would naturally expect
\bea\tdh_{\rm ppE}(f)&=&\sum_{\ell m}\tdh_{\ell m}^{\rm ppE}(f)\,,\nn\\
\tdh_{\ell m}^{\rm ppE}(f)&=&\tdh^{\rm GR}_{\ell m}\left(f\right )e^{i\beta_{\ell m}u^{b_{\ell m}}}\,,\label{eq:phase3}\eea
where both $\beta_{\ell m}$ and $b_{\ell m}$ can be different for different values of $\ell$ and $m\,$. Indeed, it has been found that \cite{Mezzasoma:2022pjb},
\be\beta_{\ell m}=\Big(\frac2m\Big)^{b/3-1}\beta_{22}\,,\quad b_{\ell m}= b_{22}\,.\label{eq.phase3b}\ee

Given an \ac{MG}, the relation between the \ac{ppE} parameters and the theory parameters can be established by calculating corrections to binary orbits \cite{Tahura:2018zuq}. For \ac{EdGB}, the leading order modification  occurs at the $-1$PN order, corresponding to $b_{22}=-7\,$, and it has been found that \cite{Yagi:2011xp}:
\bea \beta_{22}=-\frac{5\zeta}{7168}\frac{(m_1^2\tilde{s}_2-m_2^2\tilde{s}_1)^2}{M^4\eta^{18/5}}\,,\label{EdGB}\eea
where $\zeta\equiv16\pi\alpha^2/M^4\,$ \cite{Kanti:1995vq}, and $\tilde{s}_n$, $n=1,2$, is the scalar charge for the $n$th component. If the component is a black hole, we have $\tilde{s}_n\equiv2(\sqrt{1-\chi_n^2}-1+\chi_n^2)/\chi_n^2\,$. If it's a neutron star, the corresponding scalar charge is zero.

\subsection{Waveform model for the merger-ringdown stage}\label{sub:mr}

Although it is tempting to use all the available \ac{GW} data to test a given \ac{MG}, such as \ac{EdGB}, there still lacks a good waveform model that properly takes into account the corresponding \ac{MG} correction at the merger and ringdown stages, due to the difficulty with numerical relativity simulation \cite{Okounkova:2019dfo,Okounkova:2019zjf,Okounkova:2020rqw} and \ac{QNM} calculations \cite{Cardoso:2019mqo,McManus:2019ulj,Baibhav:2023clw,Bao:2019kgt,Glampedakis:2017dvb,Blazquez-Salcedo:2016enn}.

In this paper, we will follow \cite{Bonilla:2022dyt} and use the following phenomenologically motivated waveforms for the merger and ringdown stages:

\begin{itemize}
\item Zero-correction. This is the simplest case when no contribution from the merger-ringdown stage is invoked,
\bea\tdh^{\rm Zero}_{\ell m}(f)=\left\{\begin{array}{ll}
\tdh^{\mathrm{GR}}_{\ell m}(f)e^{i \beta_{\ell m} u^{b_{\ell m}}}, & f<f^{\mathrm{IM}}_{\ell m} \\
0, & f\geq f^{\mathrm{IM}}_{\ell m}
\end{array}\right.\,,\label{eq:zero}\eea
where $f^{\mathrm{IM}}_{\ell m}$ is the \ac{GW} frequency when the binary system reaches its \ac{MECO} as defined in \cite{Cabero:2016ayq}, i.e. $Mf_{\mathrm{IM}}^{22}=0.014$ and $f_{\mathrm{IM}}^{\ell m}=\frac{m}{2} f_{\mathrm{IM}}^{22}$ for IMRPhenomXHM \cite{Garcia-Quiros:2020qpx}.

\item $C^0$-correction. In this case, the correction in the merger-ringdown stage is modeled with a fixed phase,
\bea \tdh_{\ell m}^{\mathrm{C^0}}(f)=\left\{\begin{array}{ll}
\tdh_{\ell m}^{\rm GR}e^{i\beta_{\ell m} u^{b_{\ell m}}}, & f<f_{\ell m}^{\mathrm{IM}} \\
\tdh_{\ell m}^{\rm GR}e^{i\beta_{\ell m} u_{\mathrm{IM}}^{b_{\ell m}}}, & f \geq f_{\ell m}^{\mathrm{IM}}
\end{array}\right.\,,\label{eq:c0}\eea
where $u_{\mathrm{IM}}=(\pi \mathcal{M} f_{\ell m}^{\mathrm{IM}})^{1/3}\,$.

\item $C^\infty$-correction: In this case, the form of the correction in the inspiral stage is carried all the way through the merger-ringdown stages,
\bea \tdh_{\ell m}^{\mathrm{C^\infty}}(f)=\tdh_{\ell m}^{\rm GR}e^{i\beta_{\ell m} u^{b_{\ell m}}}\,.\label{eq:cinf}\eea
\end{itemize}

\section{Data source and data analysis methods}\label{sec:model}

In this section, we explain how the \ac{GW} events are selected and clarify the specifics of the statistical method used in this paper.

\subsection{Selection of \ac{GW} events}\label{sub:sour}

There are 90 \ac{GW} events in the current GWTC-3 catalog \cite{LIGOScientific:2021djp}, and using all of them to test \ac{EdGB} would be too expensive for computation, so it is preferable to choose a limited number of events that can put \ac{EdGB} under the most stringent test. As guidance for selecting the events, we would like to have reliable and strong events that have low \ac{FAR} (i.e., \ac{FAR} $<10^{-3}$ Year$^{-1}\,$) and high \ac{SNR} (i.e., \ac{SNR} $>10\,$), and we also want events that have low total mass but large mass ratio. We are thus led to the set of \ac{GW} events listed in TABLE \ref{tab:events}, guided mainly by the requirement on \ac{SNR}, total mass and mass ratio. For later use, we also list the mass, denoted as $M_*$, of the smallest black hole involved in each \ac{GW} event.\footnote{If the system is a \ac{BHB}, then $M_\ast$ is the mass of the minor component, but if the system is a \ac{NSBH}, then $M_*$ is the mass of the black hole.} It turns out that all these events have \ac{FAR} $<10^{-5}$ Year$^{-1}\,$. We will consider two possibilities for GW190814: one as a \ac{NSBH}, denoted as GW190814$^a$, and one as a \ac{BHB}, denoted as GW190814$^b$. The selected events can be roughly divided into two groups:
\begin{itemize}
\item High mass ratio events with $q\geq3\,$;
\item Low mass ratio events with $3>q\geq1\,$.
\end{itemize}
One can see that only three events belong to the high mass ratio group. For events in the high mass ratio group, constraints on \ac{EdGB} with higher harmonics contribution has been studied before \cite{Wang:2021jfc,Lyu:2022gdr}, while for events in the low mass ratio group, no higher harmonic contribution has been considered \cite{Perkins:2021mhb}.

\begin{table}
	\centering
	\caption{The list of selected \ac{GW} events. $M_*$ is the mass of the smallest black hole involved in each event.}
	\renewcommand{\arraystretch}{1.2}
	\begin{tabular}{|c|p{1.5cm}<{\centering}|p{1.5cm}<{\centering}|p{1.2cm}<{\centering}|p{1.2cm}<{\centering}|}
		\hline
		Designation & $M(\mSun)$ & $M_*(\mSun)$ & $q$ & SNR \\
		\hline
		GW190412 & 36.8  & 9.0 & 3.08  &  19.8 \\ \hline
		GW19081$4^a$ & 25.9  & 23.3 & 8.96  &  25.3 \\ \hline
		GW19081$4^b$ & 25.9  & 2.6 & 8.96  &  25.3 \\ \hline
		GW200115 & 7.4 & 5.9    & 4.1   &  11.3 \\ \hline \hline
		GW190707 & 20.1 & 7.9    & 1.53  &  13.1 \\ \hline
		GW190720 & 21.8 & 7.5   & 1.89  &  10.9 \\ \hline
		GW190728 & 20.7 & 8.0   & 1.56  &  13.1 \\ \hline
		GW190924 & 13.9 & 5.1   & 1.73  &  12.0 \\ \hline
		GW191129 & 17.5 & 6.7   & 1.6   &  13.1 \\ \hline
		GW200202 & 17.58 & 7.3  & 1.38  &  10.8 \\ \hline
	\end{tabular}
\label{tab:events}
\end{table}

\subsection{Bayesian inference}\label{sub:bayes}

The statistics are done with Bayesian inference. According to the Bayes' theorem, given the data $d$ and hypothesis $H$, the posterior probability distribution of a set of parameters $\theta$ is given by
\bea p(\theta|d, H)=\frac{p(d|\theta, H) \pi(\theta|H)}{\mathcal{Z}(d|H)}\,,\label{eq:Bayes_theorem}\eea
where $p(d|\theta, H)$ and $\pi(\theta|H)$ are the likelihood and prior, respectively. The evidence $\mathcal{Z}(d|H)$ is an overall normalization and will not be used in our calculation.

\begin{table}
\centering
\caption{\ac{GW} and theory parameters involved in this study.}
\renewcommand{\arraystretch}{1.2}
\begin{tabular}{|c|c|}
  \hline
  Symbol & Physical meaning \\
  \hline
  $m_1$             & Mass of the major component \\ \hline
  $m_2$             & Mass of the minor component \\ \hline
  $\chi_1$             & Spin of the major component \\ \hline
  $\chi_2$             & Spin of the minor component \\ \hline
  $\alpha_S$        & Right ascension of source location \\ \hline
  $\delta$          & Declination of the source location \\ \hline
  $\psi$            & Polarization angle \\ \hline
  $\iota$           & Inclination angle \\ \hline
  $\phi_{\rm ref}$  & Phase at the reference frequency \\ \hline
  $t_c$             & Coalescence time \\ \hline
  $D_L$             & Luminosity distance \\ \hline
  $\sqrt{|\alpha|}$ & Parameter from the EdGB coupling \\ \hline
\end{tabular}
\label{tab:paras}
\end{table}

The \ac{GW} and theory parameters used in this work are listed in TABLE \ref{tab:paras}. Note $\chi_1$ and $\chi_2$ are dimensionless and the spins are assumed to be aligned with the orbital angular momentum. Assuming stationary and Gaussian noise, the likelihood can be written as
\bea p(d|\theta, H) \propto \exp \Bigg[-\frac{1}{2} \sum_{j=1}^{N}(d_{j}-h_{j}|d_{j}-h_{j})\Bigg]\,,\eea
where inner product is defined as \cite{Finn:1992wt}
\bea (x|y)= 2\int_{f_{\rm low}}^{f_{\rm high}} \frac{\tilde{x}^*(f)\tilde{y}(f) + \tilde{x}(f)\tilde{y}^*(f)}{S_n(f)} df\,.\label{eq:inner_prod}\eea
Here $\td{x}^\ast$ means the complex conjugate of the Fourier component of $x$, $S_n(f)$ is the power spectral density of the detector, $f_{\rm low}$ and $f_{\rm high}$ are the frequency bounds. We assume the prior to be uniform for $m_1$, $m_2$, $\chi_1$, $\chi_2$, $\phi_{\rm ref}$ and $t_c$, the events are spatially uniformly distributed, and $\sqrt{|\alpha|}$ is uniform in $[0, 15]\ukm$.

We have used the PyCBC package \cite{Biwer:2018osg} to do the Bayesian inference, and the \ac{MCMC} sampling is done with the emcee\_pt sampler \cite{Foreman_Mackey_2013}.  We use $32$ s of data for all the events selected and set $f_{\rm low} = 20$ Hz.

\section{RESULTS}\label{sec:re}

In this section, we present the main findings of this work.

\subsection{Constraints from the inspiral stage}\label{sub:Ins}

\begin{table}
\centering
\caption{90\% constraints on $\sqrt{|\alpha|}$ using the inspiral data from selected \ac{GW} events.}
\renewcommand{\arraystretch}{1.2}
\begin{tabular}{|m{2cm}<{\centering}|m{1.5cm}<{\centering}|m{1.5cm}<{\centering}|m{1.5cm}<{\centering}|} 
  \hline
  GW event & 90\% Constr. & Existing result & Improv- ement  \\ \hline
  GW190412& 3.18 & 4.46 \cite{Wang:2021jfc} & 29\% \\ \hline
  GW190814$^a$  & 2.18 & 2.72 \cite{Lyu:2022gdr} & 20\%  \\ \hline
  GW190814$^b$  & 0.27 & 0.4 \cite{Wang:2021jfc}  0.37 \cite{Lyu:2022gdr} & 32\% \ \ \ \ \ 25\% \\ \hline
  GW200115  & 1.1 & 1.33 \cite{Lyu:2022gdr} & 17\% \\ \hline\hline
  GW190707  & 3.03 & 6.59 \cite{Perkins:2021mhb} & 54\% \\ \hline
  GW190720  & 3.74  & 6.90 \cite{Perkins:2021mhb} & 46\% \\ \hline
  GW190728  & 3.47 & 6.87 \cite{Perkins:2021mhb} & 50\% \\ \hline
  GW190924 & 2.26 & 2.98 \cite{Perkins:2021mhb} & 24\% \\ \hline
  GW191129 & 3.27 & -- & -- \\ \hline
  GW200202 & 3.79 & -- & -- \\ \hline \hline
  Comb.1 & 1.00 & --& -- \\ \hline
  Comb.2 & 0.25 & --& -- \\ \hline
  Comb.3 & 0.98 & --& -- \\ \hline
\end{tabular}
\label{tab:ins}
\end{table}

\begin{figure*}
\centering{\includegraphics[width=1.05\textwidth]{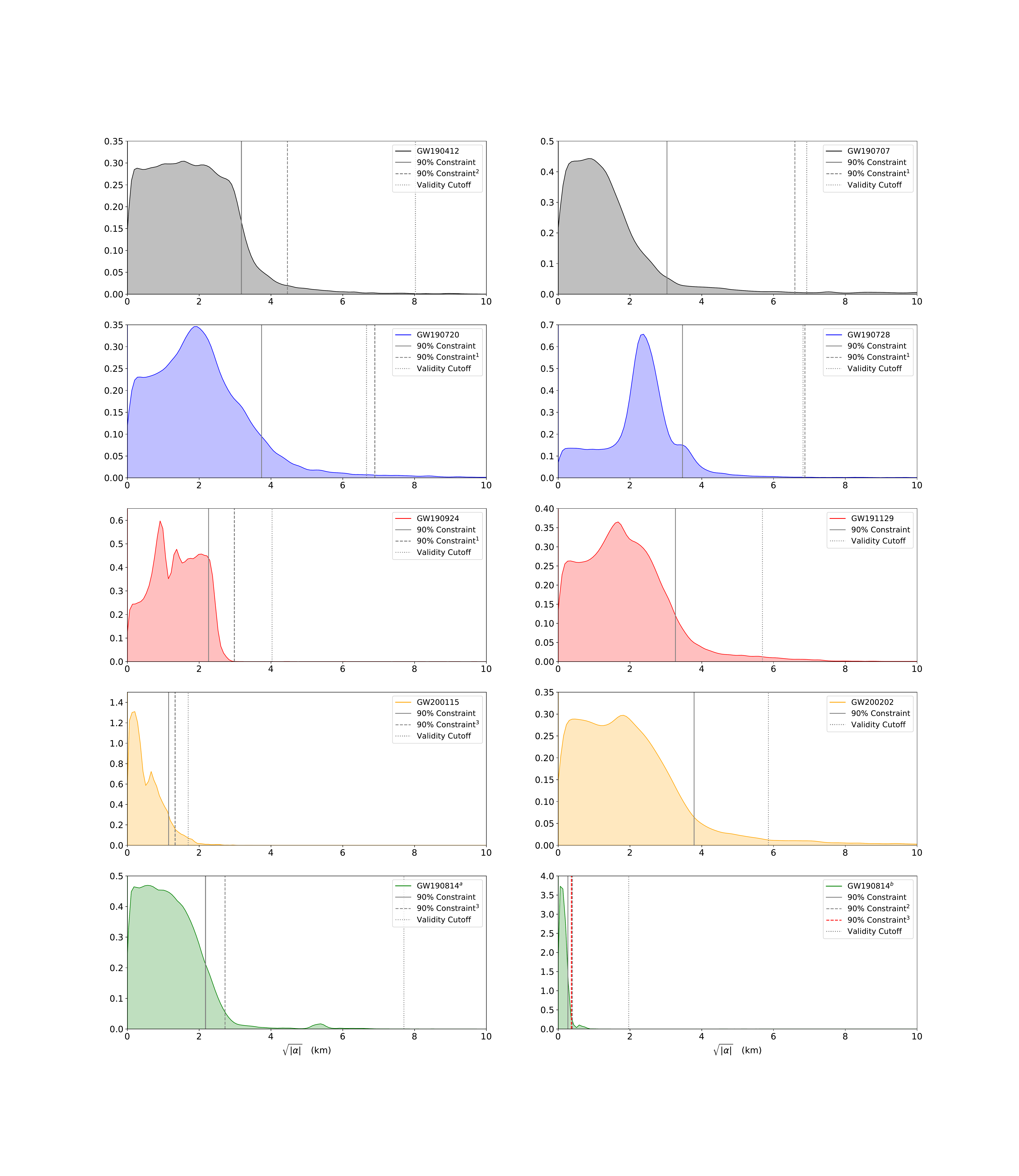}}
\caption{The cumulative posterior distributions for $\sqrt{|\alpha|}$ obtained with the inspiral data of selected \ac{GW} events. In each panel, the solid vertical line stands for the constraint on $\sqrt{|\alpha|}$ with 90$\%$ probability; the dashed vertical line stands for existing results from the literature, the superscripts 1, 2, and 3 correspond to references \cite{Perkins:2021mhb}, \cite{Wang:2021jfc} and \cite{Lyu:2022gdr}, respectively, and the corresponding values are also listed in TABLE \ref{tab:ins}; the dotted vertical line stands for the location where the weak coupling limit, characterized by $\alpha^2\lesssim\frac{m_2^4}{32 \pi}\,$ \cite{Perkins:2021mhb,Lyu:2022gdr}, is saturated, and above which the results are no longer reliable.}
\label{fig:ins}
\end{figure*}

\begin{figure}
	\centering{\includegraphics[width=0.5\textwidth]{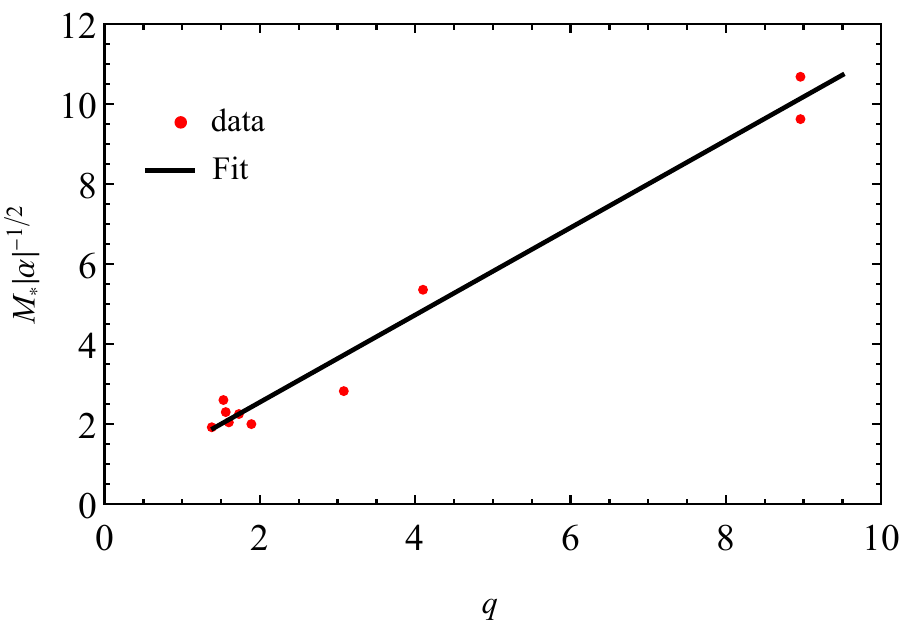}}
	\caption{The distribution of constraints from different events.}
	\label{fig:fit}
\end{figure}

Existing \ac{GW} constraints on \ac{EdGB} have been obtained only using the inspiral data. The contribution of higher harmonics has been considered in \cite{Wang:2021jfc,Lyu:2022gdr}, but a universal \ac{EdGB} correction to all harmonics has been assumed. Here we revisit the problem by using the corrected waveform model (\ref{eq:phase3}), which has been obtained in \cite{Mezzasoma:2022pjb} recently. An independent Bayesian inference has been carried out for each \ac{GW} source listed in TABLE \ref{tab:events}, except that GW190814 has been used twice, firstly as a \ac{NSBH} and then as a \ac{BHB}.

The posterior distribution of $\sqrt{|\alpha|}$ is obtained by marginalizing over all other parameters and the results are plotted in \figurename\ref{fig:ins}. The corresponding 90\% constraints are listed in TABLE \ref{tab:ins}. One can see that all the selected \ac{GW} events can constrain $\sqrt{|\alpha|}$ to better than about $4\ukm$.  The strongest constraint comes from GW190814$^b$, giving $\sqrt{|\alpha|}<0.27\ukm$, which is about $25\%$ improvement over the previous results that have taken higher harmonics into consideration \cite{Wang:2021jfc,Lyu:2022gdr}. If GW190814 is not a \ac{BHB}, then the strongest constraint comes from GW200115, giving $\sqrt{|\alpha|}<1.1\ukm$, which is about $17\%$ improvement over the existing result \cite{Lyu:2022gdr}. Except for GW190814$^b$, the strongest constraint obtained from a \ac{BHB} event comes from GW190924, which gives  $\sqrt{|\alpha|}<2.26\ukm$.

An inspection of TABLE \ref{tab:events} and TABLE \ref{tab:ins} suggests that there might be some simple approximate relation among the constraints and the parameters of different events, we are thus led to produce the plot given in FIG. \ref{fig:fit}. One can see that all the constraints found in TABLE \ref{tab:ins} are distributed not far away from the line,
\bea\frac{M_\ast}{\sqrt{|\alpha|}}\approx0.37+1.1q\,.\label{eq.fit}\eea
We do not know a reason that could lead to such a simple relation and we think more data (especially those with large mass ratios) will help clarify if there is indeed such a trend.

We also obtain combined constraints on $\sqrt{|\alpha|}$ by superimposing the posteriors of individual events \cite{LIGOScientific:2016lio}. Three combinations have been considered:
\begin{itemize}
\item Comb.1: Including all single events in TABLE \ref{tab:ins} but not GW190814$^a$ or GW190814$^b$;
\item Comb.2: Including all single events in TABLE \ref{tab:ins} but not GW190814$^a$;
\item Comb.3: Including all single events in TABLE \ref{tab:ins} but not GW190814$^b$;
\end{itemize}
One can see that the constraints can reach $1\ukm$ or better for all three combinations.

\subsection{The effect of the merger-ringdown data}\label{sub:mrr}

\begin{table}
  \centering
  \caption{Constraints on $\sqrt{|\alpha|}$ with merge-ringdown contributions. The percentage values in the lines of $C^0$ and $C^\infty$ stand for the improvement made compared to the case of Zero-correction.}
  \renewcommand{\arraystretch}{1.2}
  \begin{tabular}{|m{1.8cm}<{\centering}|m{1cm}<{\centering}|m{1cm}<{\centering}|m{1.6cm}<{\centering}|m{1.6cm}<{\centering}|}
  \hline
  \multirow{2}{*}{GW events} & \multicolumn{2}{c|}{GW190814}&\multirow{2}{*}{GW190924}& \multirow{2}{*}{GW200115} \\ \cline{2-3}
  & BHB & NSBH &  & \\ \hline
  Zero-Correction& 0.27 & 2.18 & 2.26 & 1.10 \\ \hline
  $C^0$ &  0.268 0.7\% & 2.07 5\% & 1.98 \ \ \ \ \ 12.4\% & 0.95 \ \ \ \ \ 13.6\% \\ \hline
  $C^\infty$ & 0.254  5.9\%  & 1.93  11.5\% & 1.86 \ \ \ \ \ 17.7\% & 0.87 \ \ \ \ \ 20.9\%\\ \hline
  \end{tabular}
\label{tab:mr}
\end{table}

Here we use the best events from the last subsection, GW190814$^b$, GW190924 and GW200115, all of which are characterized by containing a very low mass black hole, to study the contribution of merger-ringdown data. We also present the results on GW190814$^a$ just to make the study of this event more complete.

\begin{figure*}
\centering{\includegraphics[width=0.95\textwidth]{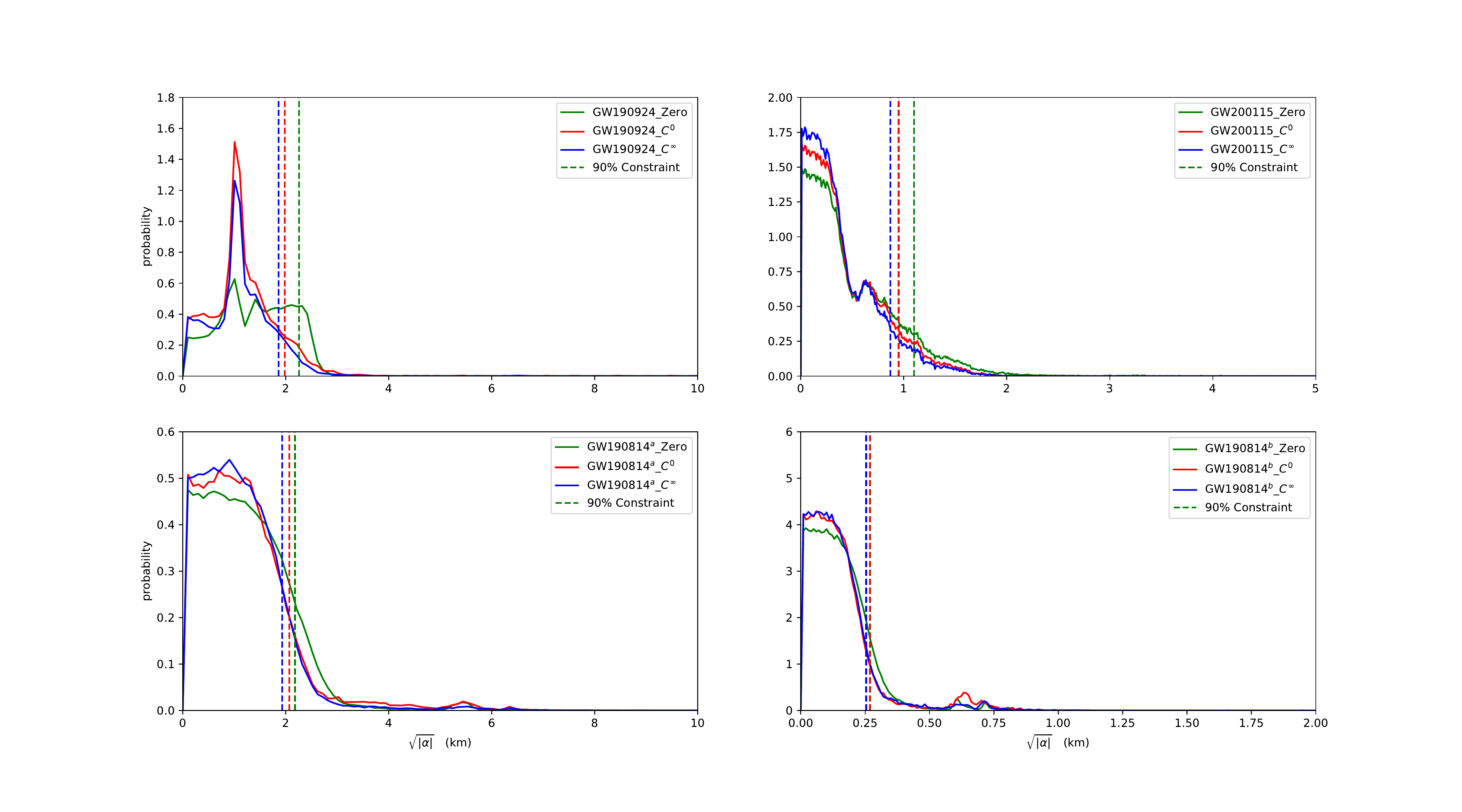}}
\caption{The cumulative posterior distributions for $\sqrt{|\alpha|}$ obtained with the full inspiral-merger-ringdown data using selected \ac{GW} events.}
\label{fig:mr}
\end{figure*}

The results are plotted in FIG.\ref{fig:mr} and the corresponding $90\%$ constraints are listed in TABLE \ref{tab:mr}. For the four cases studied, we see that the posterior distributions of $C^0$- and $C^\infty$-corrections can be appreciably different from that of Zero-correction, while the difference between $C^0$- and $C^\infty$-corrections is relatively less significant. One can also see that the merger-ringdown part can make an appreciable correction to the constraints, reaching as large as $20\%$ for the cases studied in this paper.

\section{conclusion}\label{sec:con}

In this paper, we have revisited the problem of constraining \ac{EdGB} with real \ac{GW} data. We have selected $9$ events from GWTC-3 \cite{collaboration2021gwtc3} and have paid particular attention to properly include the higher harmonics in the waveform model for the inspiral stage and have considered two schemes to include the contribution of the merger-ringdown data.

We find that different ways to include the higher harmonics can make a significant difference in the resulted constraints. For example, for some of the low mass ratio sources, such as GW190707 and GW190728, for which the higher harmonics have never been used to constrain \ac{EdGB} in previous studies, we find that the improvement can be as large as $50\%\,$. For the high mass ratio events, for which the higher harmonics have already been considered in previous studies, but by using the correct relation among the corrections to different harmonics \cite{Mezzasoma:2022pjb}, one can still get improved results. In the case of GW190412, for example, the improvement can be as large as $29\%\,$.

We also find that the most stringent constraint comes from the events that contain the least massive black holes. In particular, all constraints from the selected \ac{GW} events seem to be closely distributed by the line (\ref{eq.fit}). This is consistent with the expectation that the constraint on $\sqrt{|\alpha|}$ is at the order of the typical curvature radius of the system \cite{Berti:2015itd}. We hope more large mass ratio events could help clarify if there is indeed such a trend.

We also considered two schemes to include the merger-ringdown data in the test. By using the \ac{GW} events that pose to give the most stringent constraint on \ac{EdGB}, we find that the contribution from the merger-ringdown stage can improve the constraint by as large as about $20\%\,$.

\begin{acknowledgments}
The authors thank Yi-Fan Wang for useful discussions and Alexander Harvey Nitz for the helpful communication on the use of PyCBC. This work has been supported  by the Guangdong Basic and Applied Basic Research Foundation(Grant No.2021A1515010319), the Guangdong Major Project of Basic and Applied Basic Research (Grant No.2019B030302001), the Natural Science Foundation of China (Grants  No.12173104).
\end{acknowledgments}

\bibliographystyle{apsrev4-1}
\bibliography{reference}

\begin{thebibliography}{69}%
\makeatletter
\providecommand \@ifxundefined [1]{%
 \@ifx{#1\undefined}
}%
\providecommand \@ifnum [1]{%
 \ifnum #1\expandafter \@firstoftwo
 \else \expandafter \@secondoftwo
 \fi
}%
\providecommand \@ifx [1]{%
 \ifx #1\expandafter \@firstoftwo
 \else \expandafter \@secondoftwo
 \fi
}%
\providecommand \natexlab [1]{#1}%
\providecommand \enquote  [1]{``#1''}%
\providecommand \bibnamefont  [1]{#1}%
\providecommand \bibfnamefont [1]{#1}%
\providecommand \citenamefont [1]{#1}%
\providecommand \href@noop [0]{\@secondoftwo}%
\providecommand \href [0]{\begingroup \@sanitize@url \@href}%
\providecommand \@href[1]{\@@startlink{#1}\@@href}%
\providecommand \@@href[1]{\endgroup#1\@@endlink}%
\providecommand \@sanitize@url [0]{\catcode `\\12\catcode `\$12\catcode
  `\&12\catcode `\#12\catcode `\^12\catcode `\_12\catcode `\%12\relax}%
\providecommand \@@startlink[1]{}%
\providecommand \@@endlink[0]{}%
\providecommand \url  [0]{\begingroup\@sanitize@url \@url }%
\providecommand \@url [1]{\endgroup\@href {#1}{\urlprefix }}%
\providecommand \urlprefix  [0]{URL }%
\providecommand \Eprint [0]{\href }%
\providecommand \doibase [0]{http://dx.doi.org/}%
\providecommand \selectlanguage [0]{\@gobble}%
\providecommand \bibinfo  [0]{\@secondoftwo}%
\providecommand \bibfield  [0]{\@secondoftwo}%
\providecommand \translation [1]{[#1]}%
\providecommand \BibitemOpen [0]{}%
\providecommand \bibitemStop [0]{}%
\providecommand \bibitemNoStop [0]{.\EOS\space}%
\providecommand \EOS [0]{\spacefactor3000\relax}%
\providecommand \BibitemShut  [1]{\csname bibitem#1\endcsname}%
\let\auto@bib@innerbib\@empty
\bibitem [{\citenamefont {Abbott}\ \emph
  {et~al.}(2016{\natexlab{a}})\citenamefont {Abbott} \emph
  {et~al.}}]{LIGOScientific:2016aoc}%
  \BibitemOpen
  \bibfield  {author} {\bibinfo {author} {\bibfnamefont {B.~P.}\ \bibnamefont
  {Abbott}} \emph {et~al.} (\bibinfo {collaboration} {LIGO Scientific,
  Virgo}),\ }\href {\doibase 10.1103/PhysRevLett.116.061102} {\bibfield
  {journal} {\bibinfo  {journal} {Phys. Rev. Lett.}\ }\textbf {\bibinfo
  {volume} {116}},\ \bibinfo {pages} {061102} (\bibinfo {year}
  {2016}{\natexlab{a}})},\ \Eprint {http://arxiv.org/abs/1602.03837}
  {arXiv:1602.03837 [gr-qc]} \BibitemShut {NoStop}%
\bibitem [{\citenamefont {Abbott}\ \emph
  {et~al.}(2019{\natexlab{a}})\citenamefont {Abbott} \emph
  {et~al.}}]{LIGOScientific:2018mvr}%
  \BibitemOpen
  \bibfield  {author} {\bibinfo {author} {\bibfnamefont {B.~P.}\ \bibnamefont
  {Abbott}} \emph {et~al.} (\bibinfo {collaboration} {LIGO Scientific,
  Virgo}),\ }\href {\doibase 10.1103/PhysRevX.9.031040} {\bibfield  {journal}
  {\bibinfo  {journal} {Phys. Rev. X}\ }\textbf {\bibinfo {volume} {9}},\
  \bibinfo {pages} {031040} (\bibinfo {year} {2019}{\natexlab{a}})},\ \Eprint
  {http://arxiv.org/abs/1811.12907} {arXiv:1811.12907 [astro-ph.HE]}
  \BibitemShut {NoStop}%
\bibitem [{\citenamefont {Abbott}\ \emph
  {et~al.}(2021{\natexlab{a}})\citenamefont {Abbott} \emph
  {et~al.}}]{LIGOScientific:2020ibl}%
  \BibitemOpen
  \bibfield  {author} {\bibinfo {author} {\bibfnamefont {R.}~\bibnamefont
  {Abbott}} \emph {et~al.} (\bibinfo {collaboration} {LIGO Scientific,
  Virgo}),\ }\href {\doibase 10.1103/PhysRevX.11.021053} {\bibfield  {journal}
  {\bibinfo  {journal} {Phys. Rev. X}\ }\textbf {\bibinfo {volume} {11}},\
  \bibinfo {pages} {021053} (\bibinfo {year} {2021}{\natexlab{a}})},\ \Eprint
  {http://arxiv.org/abs/2010.14527} {arXiv:2010.14527 [gr-qc]} \BibitemShut
  {NoStop}%
\bibitem [{\citenamefont {Abbott}\ \emph
  {et~al.}(2021{\natexlab{b}})\citenamefont {Abbott} \emph
  {et~al.}}]{LIGOScientific:2021usb}%
  \BibitemOpen
  \bibfield  {author} {\bibinfo {author} {\bibfnamefont {R.}~\bibnamefont
  {Abbott}} \emph {et~al.} (\bibinfo {collaboration} {LIGO Scientific,
  VIRGO}),\ }\href@noop {} {\  (\bibinfo {year} {2021}{\natexlab{b}})},\
  \Eprint {http://arxiv.org/abs/2108.01045} {arXiv:2108.01045 [gr-qc]}
  \BibitemShut {NoStop}%
\bibitem [{\citenamefont {Abbott}\ \emph
  {et~al.}(2021{\natexlab{c}})\citenamefont {Abbott} \emph
  {et~al.}}]{LIGOScientific:2021djp}%
  \BibitemOpen
  \bibfield  {author} {\bibinfo {author} {\bibfnamefont {R.}~\bibnamefont
  {Abbott}} \emph {et~al.} (\bibinfo {collaboration} {LIGO Scientific, VIRGO,
  KAGRA}),\ }\href@noop {} {\  (\bibinfo {year} {2021}{\natexlab{c}})},\
  \Eprint {http://arxiv.org/abs/2111.03606} {arXiv:2111.03606 [gr-qc]}
  \BibitemShut {NoStop}%
\bibitem [{\citenamefont {Abbott}\ \emph
  {et~al.}(2016{\natexlab{b}})\citenamefont {Abbott} \emph
  {et~al.}}]{LIGOScientific:2016lio}%
  \BibitemOpen
  \bibfield  {author} {\bibinfo {author} {\bibfnamefont {B.~P.}\ \bibnamefont
  {Abbott}} \emph {et~al.} (\bibinfo {collaboration} {LIGO Scientific,
  Virgo}),\ }\href {\doibase 10.1103/PhysRevLett.116.221101} {\bibfield
  {journal} {\bibinfo  {journal} {Phys. Rev. Lett.}\ }\textbf {\bibinfo
  {volume} {116}},\ \bibinfo {pages} {221101} (\bibinfo {year}
  {2016}{\natexlab{b}})},\ \bibinfo {note} {[Erratum: Phys.Rev.Lett. 121,
  129902 (2018)]},\ \Eprint {http://arxiv.org/abs/1602.03841} {arXiv:1602.03841
  [gr-qc]} \BibitemShut {NoStop}%
\bibitem [{\citenamefont {Abbott}\ \emph
  {et~al.}(2019{\natexlab{b}})\citenamefont {Abbott} \emph
  {et~al.}}]{LIGOScientific:2018dkp}%
  \BibitemOpen
  \bibfield  {author} {\bibinfo {author} {\bibfnamefont {B.~P.}\ \bibnamefont
  {Abbott}} \emph {et~al.} (\bibinfo {collaboration} {LIGO Scientific,
  Virgo}),\ }\href {\doibase 10.1103/PhysRevLett.123.011102} {\bibfield
  {journal} {\bibinfo  {journal} {Phys. Rev. Lett.}\ }\textbf {\bibinfo
  {volume} {123}},\ \bibinfo {pages} {011102} (\bibinfo {year}
  {2019}{\natexlab{b}})},\ \Eprint {http://arxiv.org/abs/1811.00364}
  {arXiv:1811.00364 [gr-qc]} \BibitemShut {NoStop}%
\bibitem [{\citenamefont {Abbott}\ \emph
  {et~al.}(2019{\natexlab{c}})\citenamefont {Abbott} \emph
  {et~al.}}]{LIGOScientific:2019fpa}%
  \BibitemOpen
  \bibfield  {author} {\bibinfo {author} {\bibfnamefont {B.~P.}\ \bibnamefont
  {Abbott}} \emph {et~al.} (\bibinfo {collaboration} {LIGO Scientific,
  Virgo}),\ }\href {\doibase 10.1103/PhysRevD.100.104036} {\bibfield  {journal}
  {\bibinfo  {journal} {Phys. Rev. D}\ }\textbf {\bibinfo {volume} {100}},\
  \bibinfo {pages} {104036} (\bibinfo {year} {2019}{\natexlab{c}})},\ \Eprint
  {http://arxiv.org/abs/1903.04467} {arXiv:1903.04467 [gr-qc]} \BibitemShut
  {NoStop}%
\bibitem [{\citenamefont {Abbott}\ \emph
  {et~al.}(2021{\natexlab{d}})\citenamefont {Abbott} \emph
  {et~al.}}]{LIGOScientific:2020tif}%
  \BibitemOpen
  \bibfield  {author} {\bibinfo {author} {\bibfnamefont {R.}~\bibnamefont
  {Abbott}} \emph {et~al.} (\bibinfo {collaboration} {LIGO Scientific,
  Virgo}),\ }\href {\doibase 10.1103/PhysRevD.103.122002} {\bibfield  {journal}
  {\bibinfo  {journal} {Phys. Rev. D}\ }\textbf {\bibinfo {volume} {103}},\
  \bibinfo {pages} {122002} (\bibinfo {year} {2021}{\natexlab{d}})},\ \Eprint
  {http://arxiv.org/abs/2010.14529} {arXiv:2010.14529 [gr-qc]} \BibitemShut
  {NoStop}%
\bibitem [{\citenamefont {Abbott}\ \emph
  {et~al.}(2021{\natexlab{e}})\citenamefont {Abbott} \emph
  {et~al.}}]{LIGOScientific:2021sio}%
  \BibitemOpen
  \bibfield  {author} {\bibinfo {author} {\bibfnamefont {R.}~\bibnamefont
  {Abbott}} \emph {et~al.} (\bibinfo {collaboration} {LIGO Scientific, VIRGO,
  KAGRA}),\ }\href@noop {} {\  (\bibinfo {year} {2021}{\natexlab{e}})},\
  \Eprint {http://arxiv.org/abs/2112.06861} {arXiv:2112.06861 [gr-qc]}
  \BibitemShut {NoStop}%
\bibitem [{\citenamefont {Perkins}\ \emph {et~al.}(2021)\citenamefont
  {Perkins}, \citenamefont {Nair}, \citenamefont {Silva},\ and\ \citenamefont
  {Yunes}}]{Perkins:2021mhb}%
  \BibitemOpen
  \bibfield  {author} {\bibinfo {author} {\bibfnamefont {S.~E.}\ \bibnamefont
  {Perkins}}, \bibinfo {author} {\bibfnamefont {R.}~\bibnamefont {Nair}},
  \bibinfo {author} {\bibfnamefont {H.~O.}\ \bibnamefont {Silva}}, \ and\
  \bibinfo {author} {\bibfnamefont {N.}~\bibnamefont {Yunes}},\ }\href
  {\doibase 10.1103/PhysRevD.104.024060} {\bibfield  {journal} {\bibinfo
  {journal} {Phys. Rev. D}\ }\textbf {\bibinfo {volume} {104}},\ \bibinfo
  {pages} {024060} (\bibinfo {year} {2021})},\ \Eprint
  {http://arxiv.org/abs/2104.11189} {arXiv:2104.11189 [gr-qc]} \BibitemShut
  {NoStop}%
\bibitem [{\citenamefont {Wang}\ \emph
  {et~al.}(2021{\natexlab{a}})\citenamefont {Wang}, \citenamefont {Tang},
  \citenamefont {Li}, \citenamefont {Han},\ and\ \citenamefont
  {Fan}}]{Wang:2021jfc}%
  \BibitemOpen
  \bibfield  {author} {\bibinfo {author} {\bibfnamefont {H.-T.}\ \bibnamefont
  {Wang}}, \bibinfo {author} {\bibfnamefont {S.-P.}\ \bibnamefont {Tang}},
  \bibinfo {author} {\bibfnamefont {P.-C.}\ \bibnamefont {Li}}, \bibinfo
  {author} {\bibfnamefont {M.-Z.}\ \bibnamefont {Han}}, \ and\ \bibinfo
  {author} {\bibfnamefont {Y.-Z.}\ \bibnamefont {Fan}},\ }\href {\doibase
  10.1103/PhysRevD.104.024015} {\bibfield  {journal} {\bibinfo  {journal}
  {Phys. Rev. D}\ }\textbf {\bibinfo {volume} {104}},\ \bibinfo {pages}
  {024015} (\bibinfo {year} {2021}{\natexlab{a}})}\BibitemShut {NoStop}%
\bibitem [{\citenamefont {Niu}\ \emph {et~al.}(2021)\citenamefont {Niu},
  \citenamefont {Zhang}, \citenamefont {Wang},\ and\ \citenamefont
  {Zhao}}]{Niu:2021nic}%
  \BibitemOpen
  \bibfield  {author} {\bibinfo {author} {\bibfnamefont {R.}~\bibnamefont
  {Niu}}, \bibinfo {author} {\bibfnamefont {X.}~\bibnamefont {Zhang}}, \bibinfo
  {author} {\bibfnamefont {B.}~\bibnamefont {Wang}}, \ and\ \bibinfo {author}
  {\bibfnamefont {W.}~\bibnamefont {Zhao}},\ }\href {\doibase
  10.3847/1538-4357/ac1d4f} {\bibfield  {journal} {\bibinfo  {journal}
  {Astrophys. J.}\ }\textbf {\bibinfo {volume} {921}},\ \bibinfo {pages} {149}
  (\bibinfo {year} {2021})},\ \Eprint {http://arxiv.org/abs/2105.13644}
  {arXiv:2105.13644 [gr-qc]} \BibitemShut {NoStop}%
\bibitem [{\citenamefont {Wang}\ \emph
  {et~al.}(2021{\natexlab{b}})\citenamefont {Wang}, \citenamefont {Shao},\ and\
  \citenamefont {Liu}}]{Wang:2021ctl}%
  \BibitemOpen
  \bibfield  {author} {\bibinfo {author} {\bibfnamefont {Z.}~\bibnamefont
  {Wang}}, \bibinfo {author} {\bibfnamefont {L.}~\bibnamefont {Shao}}, \ and\
  \bibinfo {author} {\bibfnamefont {C.}~\bibnamefont {Liu}},\ }\href {\doibase
  10.3847/1538-4357/ac223c} {\bibfield  {journal} {\bibinfo  {journal}
  {Astrophys. J.}\ }\textbf {\bibinfo {volume} {921}},\ \bibinfo {pages} {158}
  (\bibinfo {year} {2021}{\natexlab{b}})},\ \Eprint
  {http://arxiv.org/abs/2108.02974} {arXiv:2108.02974 [gr-qc]} \BibitemShut
  {NoStop}%
\bibitem [{\citenamefont {Kobakhidze}\ \emph {et~al.}(2016)\citenamefont
  {Kobakhidze}, \citenamefont {Lagger},\ and\ \citenamefont
  {Manning}}]{Kobakhidze:2016cqh}%
  \BibitemOpen
  \bibfield  {author} {\bibinfo {author} {\bibfnamefont {A.}~\bibnamefont
  {Kobakhidze}}, \bibinfo {author} {\bibfnamefont {C.}~\bibnamefont {Lagger}},
  \ and\ \bibinfo {author} {\bibfnamefont {A.}~\bibnamefont {Manning}},\ }\href
  {\doibase 10.1103/PhysRevD.94.064033} {\bibfield  {journal} {\bibinfo
  {journal} {Phys. Rev. D}\ }\textbf {\bibinfo {volume} {94}},\ \bibinfo
  {pages} {064033} (\bibinfo {year} {2016})},\ \Eprint
  {http://arxiv.org/abs/1607.03776} {arXiv:1607.03776 [gr-qc]} \BibitemShut
  {NoStop}%
\bibitem [{\citenamefont {Yunes}\ \emph {et~al.}(2016)\citenamefont {Yunes},
  \citenamefont {Yagi},\ and\ \citenamefont {Pretorius}}]{Yunes:2016jcc}%
  \BibitemOpen
  \bibfield  {author} {\bibinfo {author} {\bibfnamefont {N.}~\bibnamefont
  {Yunes}}, \bibinfo {author} {\bibfnamefont {K.}~\bibnamefont {Yagi}}, \ and\
  \bibinfo {author} {\bibfnamefont {F.}~\bibnamefont {Pretorius}},\ }\href
  {\doibase 10.1103/PhysRevD.94.084002} {\bibfield  {journal} {\bibinfo
  {journal} {Phys. Rev. D}\ }\textbf {\bibinfo {volume} {94}},\ \bibinfo
  {pages} {084002} (\bibinfo {year} {2016})},\ \Eprint
  {http://arxiv.org/abs/1603.08955} {arXiv:1603.08955 [gr-qc]} \BibitemShut
  {NoStop}%
\bibitem [{\citenamefont {Ghosh}\ \emph {et~al.}(2016)\citenamefont {Ghosh}
  \emph {et~al.}}]{Ghosh:2016qgn}%
  \BibitemOpen
  \bibfield  {author} {\bibinfo {author} {\bibfnamefont {A.}~\bibnamefont
  {Ghosh}} \emph {et~al.},\ }\href {\doibase 10.1103/PhysRevD.94.021101}
  {\bibfield  {journal} {\bibinfo  {journal} {Phys. Rev. D}\ }\textbf {\bibinfo
  {volume} {94}},\ \bibinfo {pages} {021101} (\bibinfo {year} {2016})},\
  \Eprint {http://arxiv.org/abs/1602.02453} {arXiv:1602.02453 [gr-qc]}
  \BibitemShut {NoStop}%
\bibitem [{\citenamefont {Arun}\ \emph {et~al.}(2006)\citenamefont {Arun},
  \citenamefont {Iyer}, \citenamefont {Qusailah},\ and\ \citenamefont
  {Sathyaprakash}}]{Arun:2006yw}%
  \BibitemOpen
  \bibfield  {author} {\bibinfo {author} {\bibfnamefont {K.~G.}\ \bibnamefont
  {Arun}}, \bibinfo {author} {\bibfnamefont {B.~R.}\ \bibnamefont {Iyer}},
  \bibinfo {author} {\bibfnamefont {M.~S.~S.}\ \bibnamefont {Qusailah}}, \ and\
  \bibinfo {author} {\bibfnamefont {B.~S.}\ \bibnamefont {Sathyaprakash}},\
  }\href {\doibase 10.1088/0264-9381/23/9/L01} {\bibfield  {journal} {\bibinfo
  {journal} {Class. Quant. Grav.}\ }\textbf {\bibinfo {volume} {23}},\ \bibinfo
  {pages} {L37} (\bibinfo {year} {2006})},\ \Eprint
  {http://arxiv.org/abs/gr-qc/0604018} {arXiv:gr-qc/0604018} \BibitemShut
  {NoStop}%
\bibitem [{\citenamefont {Isi}\ \emph {et~al.}(2019)\citenamefont {Isi},
  \citenamefont {Giesler}, \citenamefont {Farr}, \citenamefont {Scheel},\ and\
  \citenamefont {Teukolsky}}]{Isi:2019aib}%
  \BibitemOpen
  \bibfield  {author} {\bibinfo {author} {\bibfnamefont {M.}~\bibnamefont
  {Isi}}, \bibinfo {author} {\bibfnamefont {M.}~\bibnamefont {Giesler}},
  \bibinfo {author} {\bibfnamefont {W.~M.}\ \bibnamefont {Farr}}, \bibinfo
  {author} {\bibfnamefont {M.~A.}\ \bibnamefont {Scheel}}, \ and\ \bibinfo
  {author} {\bibfnamefont {S.~A.}\ \bibnamefont {Teukolsky}},\ }\href {\doibase
  10.1103/PhysRevLett.123.111102} {\bibfield  {journal} {\bibinfo  {journal}
  {Phys. Rev. Lett.}\ }\textbf {\bibinfo {volume} {123}},\ \bibinfo {pages}
  {111102} (\bibinfo {year} {2019})},\ \Eprint
  {http://arxiv.org/abs/1905.00869} {arXiv:1905.00869 [gr-qc]} \BibitemShut
  {NoStop}%
\bibitem [{\citenamefont {Eardley}\ \emph {et~al.}(1973)\citenamefont
  {Eardley}, \citenamefont {Lee},\ and\ \citenamefont
  {Lightman}}]{Eardley:1973zuo}%
  \BibitemOpen
  \bibfield  {author} {\bibinfo {author} {\bibfnamefont {D.~M.}\ \bibnamefont
  {Eardley}}, \bibinfo {author} {\bibfnamefont {D.~L.}\ \bibnamefont {Lee}}, \
  and\ \bibinfo {author} {\bibfnamefont {A.~P.}\ \bibnamefont {Lightman}},\
  }\href {\doibase 10.1103/PhysRevD.8.3308} {\bibfield  {journal} {\bibinfo
  {journal} {Phys. Rev. D}\ }\textbf {\bibinfo {volume} {8}},\ \bibinfo {pages}
  {3308} (\bibinfo {year} {1973})}\BibitemShut {NoStop}%
\bibitem [{\citenamefont {Isi}\ \emph {et~al.}(2017)\citenamefont {Isi},
  \citenamefont {Pitkin},\ and\ \citenamefont {Weinstein}}]{Isi:2017equ}%
  \BibitemOpen
  \bibfield  {author} {\bibinfo {author} {\bibfnamefont {M.}~\bibnamefont
  {Isi}}, \bibinfo {author} {\bibfnamefont {M.}~\bibnamefont {Pitkin}}, \ and\
  \bibinfo {author} {\bibfnamefont {A.~J.}\ \bibnamefont {Weinstein}},\ }\href
  {\doibase 10.1103/PhysRevD.96.042001} {\bibfield  {journal} {\bibinfo
  {journal} {Phys. Rev. D}\ }\textbf {\bibinfo {volume} {96}},\ \bibinfo
  {pages} {042001} (\bibinfo {year} {2017})},\ \Eprint
  {http://arxiv.org/abs/1703.07530} {arXiv:1703.07530 [gr-qc]} \BibitemShut
  {NoStop}%
\bibitem [{\citenamefont {Krishnendu}\ \emph {et~al.}(2017)\citenamefont
  {Krishnendu}, \citenamefont {Arun},\ and\ \citenamefont
  {Mishra}}]{Krishnendu:2017shb}%
  \BibitemOpen
  \bibfield  {author} {\bibinfo {author} {\bibfnamefont {N.~V.}\ \bibnamefont
  {Krishnendu}}, \bibinfo {author} {\bibfnamefont {K.~G.}\ \bibnamefont
  {Arun}}, \ and\ \bibinfo {author} {\bibfnamefont {C.~K.}\ \bibnamefont
  {Mishra}},\ }\href {\doibase 10.1103/PhysRevLett.119.091101} {\bibfield
  {journal} {\bibinfo  {journal} {Phys. Rev. Lett.}\ }\textbf {\bibinfo
  {volume} {119}},\ \bibinfo {pages} {091101} (\bibinfo {year} {2017})},\
  \Eprint {http://arxiv.org/abs/1701.06318} {arXiv:1701.06318 [gr-qc]}
  \BibitemShut {NoStop}%
\bibitem [{\citenamefont {Mirshekari}\ \emph {et~al.}(2012)\citenamefont
  {Mirshekari}, \citenamefont {Yunes},\ and\ \citenamefont
  {Will}}]{Mirshekari:2011yq}%
  \BibitemOpen
  \bibfield  {author} {\bibinfo {author} {\bibfnamefont {S.}~\bibnamefont
  {Mirshekari}}, \bibinfo {author} {\bibfnamefont {N.}~\bibnamefont {Yunes}}, \
  and\ \bibinfo {author} {\bibfnamefont {C.~M.}\ \bibnamefont {Will}},\ }\href
  {\doibase 10.1103/PhysRevD.85.024041} {\bibfield  {journal} {\bibinfo
  {journal} {Phys. Rev. D}\ }\textbf {\bibinfo {volume} {85}},\ \bibinfo
  {pages} {024041} (\bibinfo {year} {2012})},\ \Eprint
  {http://arxiv.org/abs/1110.2720} {arXiv:1110.2720 [gr-qc]} \BibitemShut
  {NoStop}%
\bibitem [{\citenamefont {Vijaykumar}\ \emph {et~al.}(2021)\citenamefont
  {Vijaykumar}, \citenamefont {Kapadia},\ and\ \citenamefont
  {Ajith}}]{Vijaykumar:2020nzc}%
  \BibitemOpen
  \bibfield  {author} {\bibinfo {author} {\bibfnamefont {A.}~\bibnamefont
  {Vijaykumar}}, \bibinfo {author} {\bibfnamefont {S.~J.}\ \bibnamefont
  {Kapadia}}, \ and\ \bibinfo {author} {\bibfnamefont {P.}~\bibnamefont
  {Ajith}},\ }\href {\doibase 10.1103/PhysRevLett.126.141104} {\bibfield
  {journal} {\bibinfo  {journal} {Phys. Rev. Lett.}\ }\textbf {\bibinfo
  {volume} {126}},\ \bibinfo {pages} {141104} (\bibinfo {year} {2021})},\
  \Eprint {http://arxiv.org/abs/2003.12832} {arXiv:2003.12832 [gr-qc]}
  \BibitemShut {NoStop}%
\bibitem [{\citenamefont {Niu}\ \emph {et~al.}(2022)\citenamefont {Niu},
  \citenamefont {Zhu},\ and\ \citenamefont {Zhao}}]{Niu:2022yhr}%
  \BibitemOpen
  \bibfield  {author} {\bibinfo {author} {\bibfnamefont {R.}~\bibnamefont
  {Niu}}, \bibinfo {author} {\bibfnamefont {T.}~\bibnamefont {Zhu}}, \ and\
  \bibinfo {author} {\bibfnamefont {W.}~\bibnamefont {Zhao}},\ }\href {\doibase
  10.1088/1475-7516/2022/12/011} {\bibfield  {journal} {\bibinfo  {journal}
  {JCAP}\ }\textbf {\bibinfo {volume} {12}},\ \bibinfo {pages} {011} (\bibinfo
  {year} {2022})},\ \Eprint {http://arxiv.org/abs/2202.05092} {arXiv:2202.05092
  [gr-qc]} \BibitemShut {NoStop}%
\bibitem [{\citenamefont {Zhao}\ \emph {et~al.}(2019)\citenamefont {Zhao},
  \citenamefont {Shao}, \citenamefont {Cao},\ and\ \citenamefont
  {Ma}}]{Zhao:2019suc}%
  \BibitemOpen
  \bibfield  {author} {\bibinfo {author} {\bibfnamefont {J.}~\bibnamefont
  {Zhao}}, \bibinfo {author} {\bibfnamefont {L.}~\bibnamefont {Shao}}, \bibinfo
  {author} {\bibfnamefont {Z.}~\bibnamefont {Cao}}, \ and\ \bibinfo {author}
  {\bibfnamefont {B.-Q.}\ \bibnamefont {Ma}},\ }\href {\doibase
  10.1103/PhysRevD.100.064034} {\bibfield  {journal} {\bibinfo  {journal}
  {Phys. Rev. D}\ }\textbf {\bibinfo {volume} {100}},\ \bibinfo {pages}
  {064034} (\bibinfo {year} {2019})},\ \Eprint
  {http://arxiv.org/abs/1907.00780} {arXiv:1907.00780 [gr-qc]} \BibitemShut
  {NoStop}%
\bibitem [{\citenamefont {Okounkova}\ \emph {et~al.}(2022)\citenamefont
  {Okounkova}, \citenamefont {Farr}, \citenamefont {Isi},\ and\ \citenamefont
  {Stein}}]{Okounkova:2021xjv}%
  \BibitemOpen
  \bibfield  {author} {\bibinfo {author} {\bibfnamefont {M.}~\bibnamefont
  {Okounkova}}, \bibinfo {author} {\bibfnamefont {W.~M.}\ \bibnamefont {Farr}},
  \bibinfo {author} {\bibfnamefont {M.}~\bibnamefont {Isi}}, \ and\ \bibinfo
  {author} {\bibfnamefont {L.~C.}\ \bibnamefont {Stein}},\ }\href {\doibase
  10.1103/PhysRevD.106.044067} {\bibfield  {journal} {\bibinfo  {journal}
  {Phys. Rev. D}\ }\textbf {\bibinfo {volume} {106}},\ \bibinfo {pages}
  {044067} (\bibinfo {year} {2022})},\ \Eprint
  {http://arxiv.org/abs/2101.11153} {arXiv:2101.11153 [gr-qc]} \BibitemShut
  {NoStop}%
\bibitem [{\citenamefont {Jenks}\ \emph {et~al.}(2020)\citenamefont {Jenks},
  \citenamefont {Yagi},\ and\ \citenamefont {Alexander}}]{Jenks:2020gbt}%
  \BibitemOpen
  \bibfield  {author} {\bibinfo {author} {\bibfnamefont {L.}~\bibnamefont
  {Jenks}}, \bibinfo {author} {\bibfnamefont {K.}~\bibnamefont {Yagi}}, \ and\
  \bibinfo {author} {\bibfnamefont {S.}~\bibnamefont {Alexander}},\ }\href
  {\doibase 10.1103/PhysRevD.102.084022} {\bibfield  {journal} {\bibinfo
  {journal} {Phys. Rev. D}\ }\textbf {\bibinfo {volume} {102}},\ \bibinfo
  {pages} {084022} (\bibinfo {year} {2020})},\ \Eprint
  {http://arxiv.org/abs/2007.09714} {arXiv:2007.09714 [gr-qc]} \BibitemShut
  {NoStop}%
\bibitem [{\citenamefont {Zhu}\ \emph {et~al.}(2022)\citenamefont {Zhu},
  \citenamefont {Zhao},\ and\ \citenamefont {Wang}}]{Zhu:2022uoq}%
  \BibitemOpen
  \bibfield  {author} {\bibinfo {author} {\bibfnamefont {T.}~\bibnamefont
  {Zhu}}, \bibinfo {author} {\bibfnamefont {W.}~\bibnamefont {Zhao}}, \ and\
  \bibinfo {author} {\bibfnamefont {A.}~\bibnamefont {Wang}},\ }\href@noop {}
  {\  (\bibinfo {year} {2022})},\ \Eprint {http://arxiv.org/abs/2211.04711}
  {arXiv:2211.04711 [gr-qc]} \BibitemShut {NoStop}%
\bibitem [{\citenamefont {Wu}\ \emph {et~al.}(2022)\citenamefont {Wu},
  \citenamefont {Zhu}, \citenamefont {Niu}, \citenamefont {Zhao},\ and\
  \citenamefont {Wang}}]{Wu:2021ndf}%
  \BibitemOpen
  \bibfield  {author} {\bibinfo {author} {\bibfnamefont {Q.}~\bibnamefont
  {Wu}}, \bibinfo {author} {\bibfnamefont {T.}~\bibnamefont {Zhu}}, \bibinfo
  {author} {\bibfnamefont {R.}~\bibnamefont {Niu}}, \bibinfo {author}
  {\bibfnamefont {W.}~\bibnamefont {Zhao}}, \ and\ \bibinfo {author}
  {\bibfnamefont {A.}~\bibnamefont {Wang}},\ }\href {\doibase
  10.1103/PhysRevD.105.024035} {\bibfield  {journal} {\bibinfo  {journal}
  {Phys. Rev. D}\ }\textbf {\bibinfo {volume} {105}},\ \bibinfo {pages}
  {024035} (\bibinfo {year} {2022})},\ \Eprint
  {http://arxiv.org/abs/2110.13870} {arXiv:2110.13870 [gr-qc]} \BibitemShut
  {NoStop}%
\bibitem [{\citenamefont {Wang}\ \emph {et~al.}(2022)\citenamefont {Wang},
  \citenamefont {Brown}, \citenamefont {Shao},\ and\ \citenamefont
  {Zhao}}]{Wang:2021gqm}%
  \BibitemOpen
  \bibfield  {author} {\bibinfo {author} {\bibfnamefont {Y.-F.}\ \bibnamefont
  {Wang}}, \bibinfo {author} {\bibfnamefont {S.~M.}\ \bibnamefont {Brown}},
  \bibinfo {author} {\bibfnamefont {L.}~\bibnamefont {Shao}}, \ and\ \bibinfo
  {author} {\bibfnamefont {W.}~\bibnamefont {Zhao}},\ }\href {\doibase
  10.1103/PhysRevD.106.084005} {\bibfield  {journal} {\bibinfo  {journal}
  {Phys. Rev. D}\ }\textbf {\bibinfo {volume} {106}},\ \bibinfo {pages}
  {084005} (\bibinfo {year} {2022})},\ \Eprint
  {http://arxiv.org/abs/2109.09718} {arXiv:2109.09718 [astro-ph.HE]}
  \BibitemShut {NoStop}%
\bibitem [{\citenamefont {Wang}\ \emph
  {et~al.}(2021{\natexlab{c}})\citenamefont {Wang}, \citenamefont {Niu},
  \citenamefont {Zhu},\ and\ \citenamefont {Zhao}}]{Wang:2020cub}%
  \BibitemOpen
  \bibfield  {author} {\bibinfo {author} {\bibfnamefont {Y.-F.}\ \bibnamefont
  {Wang}}, \bibinfo {author} {\bibfnamefont {R.}~\bibnamefont {Niu}}, \bibinfo
  {author} {\bibfnamefont {T.}~\bibnamefont {Zhu}}, \ and\ \bibinfo {author}
  {\bibfnamefont {W.}~\bibnamefont {Zhao}},\ }\href {\doibase
  10.3847/1538-4357/abd7a6} {\bibfield  {journal} {\bibinfo  {journal}
  {Astrophys. J.}\ }\textbf {\bibinfo {volume} {908}},\ \bibinfo {pages} {58}
  (\bibinfo {year} {2021}{\natexlab{c}})},\ \Eprint
  {http://arxiv.org/abs/2002.05668} {arXiv:2002.05668 [gr-qc]} \BibitemShut
  {NoStop}%
\bibitem [{\citenamefont {Haegel}\ \emph {et~al.}(2022)\citenamefont {Haegel},
  \citenamefont {O'Neal-Ault}, \citenamefont {Bailey}, \citenamefont {Tasson},
  \citenamefont {Bloom},\ and\ \citenamefont {Shao}}]{Haegel:2022ymk}%
  \BibitemOpen
  \bibfield  {author} {\bibinfo {author} {\bibfnamefont {L.}~\bibnamefont
  {Haegel}}, \bibinfo {author} {\bibfnamefont {K.}~\bibnamefont {O'Neal-Ault}},
  \bibinfo {author} {\bibfnamefont {Q.~G.}\ \bibnamefont {Bailey}}, \bibinfo
  {author} {\bibfnamefont {J.~D.}\ \bibnamefont {Tasson}}, \bibinfo {author}
  {\bibfnamefont {M.}~\bibnamefont {Bloom}}, \ and\ \bibinfo {author}
  {\bibfnamefont {L.}~\bibnamefont {Shao}},\ }\href@noop {} {\  (\bibinfo
  {year} {2022})},\ \Eprint {http://arxiv.org/abs/2210.04481} {arXiv:2210.04481
  [gr-qc]} \BibitemShut {NoStop}%
\bibitem [{\citenamefont {Gong}\ \emph {et~al.}(2022)\citenamefont {Gong},
  \citenamefont {Zhu}, \citenamefont {Niu}, \citenamefont {Wu}, \citenamefont
  {Cui}, \citenamefont {Zhang}, \citenamefont {Zhao},\ and\ \citenamefont
  {Wang}}]{Gong:2021jgg}%
  \BibitemOpen
  \bibfield  {author} {\bibinfo {author} {\bibfnamefont {C.}~\bibnamefont
  {Gong}}, \bibinfo {author} {\bibfnamefont {T.}~\bibnamefont {Zhu}}, \bibinfo
  {author} {\bibfnamefont {R.}~\bibnamefont {Niu}}, \bibinfo {author}
  {\bibfnamefont {Q.}~\bibnamefont {Wu}}, \bibinfo {author} {\bibfnamefont
  {J.-L.}\ \bibnamefont {Cui}}, \bibinfo {author} {\bibfnamefont
  {X.}~\bibnamefont {Zhang}}, \bibinfo {author} {\bibfnamefont
  {W.}~\bibnamefont {Zhao}}, \ and\ \bibinfo {author} {\bibfnamefont
  {A.}~\bibnamefont {Wang}},\ }\href {\doibase 10.1103/PhysRevD.105.044034}
  {\bibfield  {journal} {\bibinfo  {journal} {Phys. Rev. D}\ }\textbf {\bibinfo
  {volume} {105}},\ \bibinfo {pages} {044034} (\bibinfo {year} {2022})},\
  \Eprint {http://arxiv.org/abs/2112.06446} {arXiv:2112.06446 [gr-qc]}
  \BibitemShut {NoStop}%
\bibitem [{\citenamefont {Du}\ \emph {et~al.}(2021)\citenamefont {Du},
  \citenamefont {Tahura}, \citenamefont {Vaman},\ and\ \citenamefont
  {Yagi}}]{Du:2020rlx}%
  \BibitemOpen
  \bibfield  {author} {\bibinfo {author} {\bibfnamefont {Y.}~\bibnamefont
  {Du}}, \bibinfo {author} {\bibfnamefont {S.}~\bibnamefont {Tahura}}, \bibinfo
  {author} {\bibfnamefont {D.}~\bibnamefont {Vaman}}, \ and\ \bibinfo {author}
  {\bibfnamefont {K.}~\bibnamefont {Yagi}},\ }\href {\doibase
  10.1103/PhysRevD.103.044031} {\bibfield  {journal} {\bibinfo  {journal}
  {Phys. Rev. D}\ }\textbf {\bibinfo {volume} {103}},\ \bibinfo {pages}
  {044031} (\bibinfo {year} {2021})},\ \Eprint
  {http://arxiv.org/abs/2004.03051} {arXiv:2004.03051 [gr-qc]} \BibitemShut
  {NoStop}%
\bibitem [{\citenamefont {Kanti}\ \emph {et~al.}(1996)\citenamefont {Kanti},
  \citenamefont {Mavromatos}, \citenamefont {Rizos}, \citenamefont {Tamvakis},\
  and\ \citenamefont {Winstanley}}]{Kanti:1995vq}%
  \BibitemOpen
  \bibfield  {author} {\bibinfo {author} {\bibfnamefont {P.}~\bibnamefont
  {Kanti}}, \bibinfo {author} {\bibfnamefont {N.~E.}\ \bibnamefont
  {Mavromatos}}, \bibinfo {author} {\bibfnamefont {J.}~\bibnamefont {Rizos}},
  \bibinfo {author} {\bibfnamefont {K.}~\bibnamefont {Tamvakis}}, \ and\
  \bibinfo {author} {\bibfnamefont {E.}~\bibnamefont {Winstanley}},\ }\href
  {\doibase 10.1103/PhysRevD.54.5049} {\bibfield  {journal} {\bibinfo
  {journal} {Phys. Rev. D}\ }\textbf {\bibinfo {volume} {54}},\ \bibinfo
  {pages} {5049} (\bibinfo {year} {1996})},\ \Eprint
  {http://arxiv.org/abs/hep-th/9511071} {arXiv:hep-th/9511071} \BibitemShut
  {NoStop}%
\bibitem [{\citenamefont {Torii}\ \emph {et~al.}(1997)\citenamefont {Torii},
  \citenamefont {Yajima},\ and\ \citenamefont {Maeda}}]{Torii:1996yi}%
  \BibitemOpen
  \bibfield  {author} {\bibinfo {author} {\bibfnamefont {T.}~\bibnamefont
  {Torii}}, \bibinfo {author} {\bibfnamefont {H.}~\bibnamefont {Yajima}}, \
  and\ \bibinfo {author} {\bibfnamefont {K.-i.}\ \bibnamefont {Maeda}},\ }\href
  {\doibase 10.1103/PhysRevD.55.739} {\bibfield  {journal} {\bibinfo  {journal}
  {Phys. Rev. D}\ }\textbf {\bibinfo {volume} {55}},\ \bibinfo {pages} {739}
  (\bibinfo {year} {1997})},\ \Eprint {http://arxiv.org/abs/gr-qc/9606034}
  {arXiv:gr-qc/9606034} \BibitemShut {NoStop}%
\bibitem [{\citenamefont {Nojiri}\ \emph {et~al.}(2005)\citenamefont {Nojiri},
  \citenamefont {Odintsov},\ and\ \citenamefont {Sasaki}}]{Nojiri:2005vv}%
  \BibitemOpen
  \bibfield  {author} {\bibinfo {author} {\bibfnamefont {S.}~\bibnamefont
  {Nojiri}}, \bibinfo {author} {\bibfnamefont {S.~D.}\ \bibnamefont
  {Odintsov}}, \ and\ \bibinfo {author} {\bibfnamefont {M.}~\bibnamefont
  {Sasaki}},\ }\href {\doibase 10.1103/PhysRevD.71.123509} {\bibfield
  {journal} {\bibinfo  {journal} {Phys. Rev. D}\ }\textbf {\bibinfo {volume}
  {71}},\ \bibinfo {pages} {123509} (\bibinfo {year} {2005})},\ \Eprint
  {http://arxiv.org/abs/hep-th/0504052} {arXiv:hep-th/0504052} \BibitemShut
  {NoStop}%
\bibitem [{\citenamefont {Yagi}(2012)}]{Yagi:2012gp}%
  \BibitemOpen
  \bibfield  {author} {\bibinfo {author} {\bibfnamefont {K.}~\bibnamefont
  {Yagi}},\ }\href {\doibase 10.1103/PhysRevD.86.081504} {\bibfield  {journal}
  {\bibinfo  {journal} {Phys. Rev. D}\ }\textbf {\bibinfo {volume} {86}},\
  \bibinfo {pages} {081504} (\bibinfo {year} {2012})},\ \Eprint
  {http://arxiv.org/abs/1204.4524} {arXiv:1204.4524 [gr-qc]} \BibitemShut
  {NoStop}%
\bibitem [{\citenamefont {Berti}\ \emph {et~al.}(2015)\citenamefont {Berti}
  \emph {et~al.}}]{Berti:2015itd}%
  \BibitemOpen
  \bibfield  {author} {\bibinfo {author} {\bibfnamefont {E.}~\bibnamefont
  {Berti}} \emph {et~al.},\ }\href {\doibase 10.1088/0264-9381/32/24/243001}
  {\bibfield  {journal} {\bibinfo  {journal} {Class. Quant. Grav.}\ }\textbf
  {\bibinfo {volume} {32}},\ \bibinfo {pages} {243001} (\bibinfo {year}
  {2015})},\ \Eprint {http://arxiv.org/abs/1501.07274} {arXiv:1501.07274
  [gr-qc]} \BibitemShut {NoStop}%
\bibitem [{\citenamefont {Yagi}\ \emph {et~al.}(2012)\citenamefont {Yagi},
  \citenamefont {Stein}, \citenamefont {Yunes},\ and\ \citenamefont
  {Tanaka}}]{Yagi:2011xp}%
  \BibitemOpen
  \bibfield  {author} {\bibinfo {author} {\bibfnamefont {K.}~\bibnamefont
  {Yagi}}, \bibinfo {author} {\bibfnamefont {L.~C.}\ \bibnamefont {Stein}},
  \bibinfo {author} {\bibfnamefont {N.}~\bibnamefont {Yunes}}, \ and\ \bibinfo
  {author} {\bibfnamefont {T.}~\bibnamefont {Tanaka}},\ }\href {\doibase
  10.1103/PhysRevD.85.064022} {\bibfield  {journal} {\bibinfo  {journal} {Phys.
  Rev. D}\ }\textbf {\bibinfo {volume} {85}},\ \bibinfo {pages} {064022}
  (\bibinfo {year} {2012})},\ \bibinfo {note} {[Erratum: Phys.Rev.D 93, 029902
  (2016)]},\ \Eprint {http://arxiv.org/abs/1110.5950} {arXiv:1110.5950 [gr-qc]}
  \BibitemShut {NoStop}%
\bibitem [{\citenamefont {Tahura}\ \emph {et~al.}(2019)\citenamefont {Tahura},
  \citenamefont {Yagi},\ and\ \citenamefont {Carson}}]{Tahura:2019dgr}%
  \BibitemOpen
  \bibfield  {author} {\bibinfo {author} {\bibfnamefont {S.}~\bibnamefont
  {Tahura}}, \bibinfo {author} {\bibfnamefont {K.}~\bibnamefont {Yagi}}, \ and\
  \bibinfo {author} {\bibfnamefont {Z.}~\bibnamefont {Carson}},\ }\href
  {\doibase 10.1103/PhysRevD.100.104001} {\bibfield  {journal} {\bibinfo
  {journal} {Phys. Rev. D}\ }\textbf {\bibinfo {volume} {100}},\ \bibinfo
  {pages} {104001} (\bibinfo {year} {2019})},\ \Eprint
  {http://arxiv.org/abs/1907.10059} {arXiv:1907.10059 [gr-qc]} \BibitemShut
  {NoStop}%
\bibitem [{\citenamefont {Nair}\ \emph {et~al.}(2019)\citenamefont {Nair},
  \citenamefont {Perkins}, \citenamefont {Silva},\ and\ \citenamefont
  {Yunes}}]{Nair:2019iur}%
  \BibitemOpen
  \bibfield  {author} {\bibinfo {author} {\bibfnamefont {R.}~\bibnamefont
  {Nair}}, \bibinfo {author} {\bibfnamefont {S.}~\bibnamefont {Perkins}},
  \bibinfo {author} {\bibfnamefont {H.~O.}\ \bibnamefont {Silva}}, \ and\
  \bibinfo {author} {\bibfnamefont {N.}~\bibnamefont {Yunes}},\ }\href
  {\doibase 10.1103/PhysRevLett.123.191101} {\bibfield  {journal} {\bibinfo
  {journal} {Phys. Rev. Lett.}\ }\textbf {\bibinfo {volume} {123}},\ \bibinfo
  {pages} {191101} (\bibinfo {year} {2019})},\ \Eprint
  {http://arxiv.org/abs/1905.00870} {arXiv:1905.00870 [gr-qc]} \BibitemShut
  {NoStop}%
\bibitem [{\citenamefont {Shi}\ \emph {et~al.}(2022)\citenamefont {Shi},
  \citenamefont {Ji}, \citenamefont {Zhang},\ and\ \citenamefont
  {Mei}}]{Shi:2022qno}%
  \BibitemOpen
  \bibfield  {author} {\bibinfo {author} {\bibfnamefont {C.}~\bibnamefont
  {Shi}}, \bibinfo {author} {\bibfnamefont {M.}~\bibnamefont {Ji}}, \bibinfo
  {author} {\bibfnamefont {J.-d.}\ \bibnamefont {Zhang}}, \ and\ \bibinfo
  {author} {\bibfnamefont {J.}~\bibnamefont {Mei}},\ }\href@noop {} {\
  (\bibinfo {year} {2022})},\ \Eprint {http://arxiv.org/abs/2210.13006}
  {arXiv:2210.13006 [gr-qc]} \BibitemShut {NoStop}%
\bibitem [{\citenamefont {Tahura}\ and\ \citenamefont
  {Yagi}(2018)}]{Tahura:2018zuq}%
  \BibitemOpen
  \bibfield  {author} {\bibinfo {author} {\bibfnamefont {S.}~\bibnamefont
  {Tahura}}\ and\ \bibinfo {author} {\bibfnamefont {K.}~\bibnamefont {Yagi}},\
  }\href {\doibase 10.1103/PhysRevD.98.084042} {\bibfield  {journal} {\bibinfo
  {journal} {Phys. Rev. D}\ }\textbf {\bibinfo {volume} {98}},\ \bibinfo
  {pages} {084042} (\bibinfo {year} {2018})},\ \bibinfo {note} {[Erratum:
  Phys.Rev.D 101, 109902 (2020)]},\ \Eprint {http://arxiv.org/abs/1809.00259}
  {arXiv:1809.00259 [gr-qc]} \BibitemShut {NoStop}%
\bibitem [{\citenamefont {Lyu}\ \emph {et~al.}(2022)\citenamefont {Lyu},
  \citenamefont {Jiang},\ and\ \citenamefont {Yagi}}]{Lyu:2022gdr}%
  \BibitemOpen
  \bibfield  {author} {\bibinfo {author} {\bibfnamefont {Z.}~\bibnamefont
  {Lyu}}, \bibinfo {author} {\bibfnamefont {N.}~\bibnamefont {Jiang}}, \ and\
  \bibinfo {author} {\bibfnamefont {K.}~\bibnamefont {Yagi}},\ }\href {\doibase
  10.1103/PhysRevD.105.064001} {\bibfield  {journal} {\bibinfo  {journal}
  {Phys. Rev. D}\ }\textbf {\bibinfo {volume} {105}},\ \bibinfo {pages}
  {064001} (\bibinfo {year} {2022})},\ \Eprint
  {http://arxiv.org/abs/2201.02543} {arXiv:2201.02543 [gr-qc]} \BibitemShut
  {NoStop}%
\bibitem [{\citenamefont {Mezzasoma}\ and\ \citenamefont
  {Yunes}(2022)}]{Mezzasoma:2022pjb}%
  \BibitemOpen
  \bibfield  {author} {\bibinfo {author} {\bibfnamefont {S.}~\bibnamefont
  {Mezzasoma}}\ and\ \bibinfo {author} {\bibfnamefont {N.}~\bibnamefont
  {Yunes}},\ }\href {\doibase 10.1103/PhysRevD.106.024026} {\bibfield
  {journal} {\bibinfo  {journal} {Phys. Rev. D}\ }\textbf {\bibinfo {volume}
  {106}},\ \bibinfo {pages} {024026} (\bibinfo {year} {2022})},\ \Eprint
  {http://arxiv.org/abs/2203.15934} {arXiv:2203.15934 [gr-qc]} \BibitemShut
  {NoStop}%
\bibitem [{\citenamefont {Okounkova}\ \emph {et~al.}(2019)\citenamefont
  {Okounkova}, \citenamefont {Stein}, \citenamefont {Scheel},\ and\
  \citenamefont {Teukolsky}}]{Okounkova:2019dfo}%
  \BibitemOpen
  \bibfield  {author} {\bibinfo {author} {\bibfnamefont {M.}~\bibnamefont
  {Okounkova}}, \bibinfo {author} {\bibfnamefont {L.~C.}\ \bibnamefont
  {Stein}}, \bibinfo {author} {\bibfnamefont {M.~A.}\ \bibnamefont {Scheel}}, \
  and\ \bibinfo {author} {\bibfnamefont {S.~A.}\ \bibnamefont {Teukolsky}},\
  }\href {\doibase 10.1103/PhysRevD.100.104026} {\bibfield  {journal} {\bibinfo
   {journal} {Phys. Rev. D}\ }\textbf {\bibinfo {volume} {100}},\ \bibinfo
  {pages} {104026} (\bibinfo {year} {2019})},\ \Eprint
  {http://arxiv.org/abs/1906.08789} {arXiv:1906.08789 [gr-qc]} \BibitemShut
  {NoStop}%
\bibitem [{\citenamefont {Okounkova}\ \emph {et~al.}(2020)\citenamefont
  {Okounkova}, \citenamefont {Stein}, \citenamefont {Moxon}, \citenamefont
  {Scheel},\ and\ \citenamefont {Teukolsky}}]{Okounkova:2019zjf}%
  \BibitemOpen
  \bibfield  {author} {\bibinfo {author} {\bibfnamefont {M.}~\bibnamefont
  {Okounkova}}, \bibinfo {author} {\bibfnamefont {L.~C.}\ \bibnamefont
  {Stein}}, \bibinfo {author} {\bibfnamefont {J.}~\bibnamefont {Moxon}},
  \bibinfo {author} {\bibfnamefont {M.~A.}\ \bibnamefont {Scheel}}, \ and\
  \bibinfo {author} {\bibfnamefont {S.~A.}\ \bibnamefont {Teukolsky}},\ }\href
  {\doibase 10.1103/PhysRevD.101.104016} {\bibfield  {journal} {\bibinfo
  {journal} {Phys. Rev. D}\ }\textbf {\bibinfo {volume} {101}},\ \bibinfo
  {pages} {104016} (\bibinfo {year} {2020})},\ \Eprint
  {http://arxiv.org/abs/1911.02588} {arXiv:1911.02588 [gr-qc]} \BibitemShut
  {NoStop}%
\bibitem [{\citenamefont {Okounkova}(2020)}]{Okounkova:2020rqw}%
  \BibitemOpen
  \bibfield  {author} {\bibinfo {author} {\bibfnamefont {M.}~\bibnamefont
  {Okounkova}},\ }\href {\doibase 10.1103/PhysRevD.102.084046} {\bibfield
  {journal} {\bibinfo  {journal} {Phys. Rev. D}\ }\textbf {\bibinfo {volume}
  {102}},\ \bibinfo {pages} {084046} (\bibinfo {year} {2020})},\ \Eprint
  {http://arxiv.org/abs/2001.03571} {arXiv:2001.03571 [gr-qc]} \BibitemShut
  {NoStop}%
\bibitem [{\citenamefont {Cardoso}\ \emph {et~al.}(2019)\citenamefont
  {Cardoso}, \citenamefont {Kimura}, \citenamefont {Maselli}, \citenamefont
  {Berti}, \citenamefont {Macedo},\ and\ \citenamefont
  {McManus}}]{Cardoso:2019mqo}%
  \BibitemOpen
  \bibfield  {author} {\bibinfo {author} {\bibfnamefont {V.}~\bibnamefont
  {Cardoso}}, \bibinfo {author} {\bibfnamefont {M.}~\bibnamefont {Kimura}},
  \bibinfo {author} {\bibfnamefont {A.}~\bibnamefont {Maselli}}, \bibinfo
  {author} {\bibfnamefont {E.}~\bibnamefont {Berti}}, \bibinfo {author}
  {\bibfnamefont {C.~F.~B.}\ \bibnamefont {Macedo}}, \ and\ \bibinfo {author}
  {\bibfnamefont {R.}~\bibnamefont {McManus}},\ }\href {\doibase
  10.1103/PhysRevD.99.104077} {\bibfield  {journal} {\bibinfo  {journal} {Phys.
  Rev. D}\ }\textbf {\bibinfo {volume} {99}},\ \bibinfo {pages} {104077}
  (\bibinfo {year} {2019})},\ \Eprint {http://arxiv.org/abs/1901.01265}
  {arXiv:1901.01265 [gr-qc]} \BibitemShut {NoStop}%
\bibitem [{\citenamefont {McManus}\ \emph {et~al.}(2019)\citenamefont
  {McManus}, \citenamefont {Berti}, \citenamefont {Macedo}, \citenamefont
  {Kimura}, \citenamefont {Maselli},\ and\ \citenamefont
  {Cardoso}}]{McManus:2019ulj}%
  \BibitemOpen
  \bibfield  {author} {\bibinfo {author} {\bibfnamefont {R.}~\bibnamefont
  {McManus}}, \bibinfo {author} {\bibfnamefont {E.}~\bibnamefont {Berti}},
  \bibinfo {author} {\bibfnamefont {C.~F.~B.}\ \bibnamefont {Macedo}}, \bibinfo
  {author} {\bibfnamefont {M.}~\bibnamefont {Kimura}}, \bibinfo {author}
  {\bibfnamefont {A.}~\bibnamefont {Maselli}}, \ and\ \bibinfo {author}
  {\bibfnamefont {V.}~\bibnamefont {Cardoso}},\ }\href {\doibase
  10.1103/PhysRevD.100.044061} {\bibfield  {journal} {\bibinfo  {journal}
  {Phys. Rev. D}\ }\textbf {\bibinfo {volume} {100}},\ \bibinfo {pages}
  {044061} (\bibinfo {year} {2019})},\ \Eprint
  {http://arxiv.org/abs/1906.05155} {arXiv:1906.05155 [gr-qc]} \BibitemShut
  {NoStop}%
\bibitem [{\citenamefont {Baibhav}\ \emph {et~al.}(2023)\citenamefont
  {Baibhav}, \citenamefont {Cheung}, \citenamefont {Berti}, \citenamefont
  {Cardoso}, \citenamefont {Carullo}, \citenamefont {Cotesta}, \citenamefont
  {Del~Pozzo},\ and\ \citenamefont {Duque}}]{Baibhav:2023clw}%
  \BibitemOpen
  \bibfield  {author} {\bibinfo {author} {\bibfnamefont {V.}~\bibnamefont
  {Baibhav}}, \bibinfo {author} {\bibfnamefont {M.~H.-Y.}\ \bibnamefont
  {Cheung}}, \bibinfo {author} {\bibfnamefont {E.}~\bibnamefont {Berti}},
  \bibinfo {author} {\bibfnamefont {V.}~\bibnamefont {Cardoso}}, \bibinfo
  {author} {\bibfnamefont {G.}~\bibnamefont {Carullo}}, \bibinfo {author}
  {\bibfnamefont {R.}~\bibnamefont {Cotesta}}, \bibinfo {author} {\bibfnamefont
  {W.}~\bibnamefont {Del~Pozzo}}, \ and\ \bibinfo {author} {\bibfnamefont
  {F.}~\bibnamefont {Duque}},\ }\href@noop {} {\  (\bibinfo {year} {2023})},\
  \Eprint {http://arxiv.org/abs/2302.03050} {arXiv:2302.03050 [gr-qc]}
  \BibitemShut {NoStop}%
\bibitem [{\citenamefont {Bao}\ \emph {et~al.}(2019)\citenamefont {Bao},
  \citenamefont {Shi}, \citenamefont {Wang}, \citenamefont {Zhang},
  \citenamefont {Hu}, \citenamefont {Mei},\ and\ \citenamefont
  {Luo}}]{Bao:2019kgt}%
  \BibitemOpen
  \bibfield  {author} {\bibinfo {author} {\bibfnamefont {J.}~\bibnamefont
  {Bao}}, \bibinfo {author} {\bibfnamefont {C.}~\bibnamefont {Shi}}, \bibinfo
  {author} {\bibfnamefont {H.}~\bibnamefont {Wang}}, \bibinfo {author}
  {\bibfnamefont {J.-d.}\ \bibnamefont {Zhang}}, \bibinfo {author}
  {\bibfnamefont {Y.}~\bibnamefont {Hu}}, \bibinfo {author} {\bibfnamefont
  {J.}~\bibnamefont {Mei}}, \ and\ \bibinfo {author} {\bibfnamefont
  {J.}~\bibnamefont {Luo}},\ }\href {\doibase 10.1103/PhysRevD.100.084024}
  {\bibfield  {journal} {\bibinfo  {journal} {Phys. Rev. D}\ }\textbf {\bibinfo
  {volume} {100}},\ \bibinfo {pages} {084024} (\bibinfo {year} {2019})},\
  \Eprint {http://arxiv.org/abs/1905.11674} {arXiv:1905.11674 [gr-qc]}
  \BibitemShut {NoStop}%
\bibitem [{\citenamefont {Glampedakis}\ \emph {et~al.}(2017)\citenamefont
  {Glampedakis}, \citenamefont {Pappas}, \citenamefont {Silva},\ and\
  \citenamefont {Berti}}]{Glampedakis:2017dvb}%
  \BibitemOpen
  \bibfield  {author} {\bibinfo {author} {\bibfnamefont {K.}~\bibnamefont
  {Glampedakis}}, \bibinfo {author} {\bibfnamefont {G.}~\bibnamefont {Pappas}},
  \bibinfo {author} {\bibfnamefont {H.~O.}\ \bibnamefont {Silva}}, \ and\
  \bibinfo {author} {\bibfnamefont {E.}~\bibnamefont {Berti}},\ }\href
  {\doibase 10.1103/PhysRevD.96.064054} {\bibfield  {journal} {\bibinfo
  {journal} {Phys. Rev. D}\ }\textbf {\bibinfo {volume} {96}},\ \bibinfo
  {pages} {064054} (\bibinfo {year} {2017})},\ \Eprint
  {http://arxiv.org/abs/1706.07658} {arXiv:1706.07658 [gr-qc]} \BibitemShut
  {NoStop}%
\bibitem [{\citenamefont {Bl\'azquez-Salcedo}\ \emph
  {et~al.}(2016)\citenamefont {Bl\'azquez-Salcedo}, \citenamefont {Macedo},
  \citenamefont {Cardoso}, \citenamefont {Ferrari}, \citenamefont {Gualtieri},
  \citenamefont {Khoo}, \citenamefont {Kunz},\ and\ \citenamefont
  {Pani}}]{Blazquez-Salcedo:2016enn}%
  \BibitemOpen
  \bibfield  {author} {\bibinfo {author} {\bibfnamefont {J.~L.}\ \bibnamefont
  {Bl\'azquez-Salcedo}}, \bibinfo {author} {\bibfnamefont {C.~F.~B.}\
  \bibnamefont {Macedo}}, \bibinfo {author} {\bibfnamefont {V.}~\bibnamefont
  {Cardoso}}, \bibinfo {author} {\bibfnamefont {V.}~\bibnamefont {Ferrari}},
  \bibinfo {author} {\bibfnamefont {L.}~\bibnamefont {Gualtieri}}, \bibinfo
  {author} {\bibfnamefont {F.~S.}\ \bibnamefont {Khoo}}, \bibinfo {author}
  {\bibfnamefont {J.}~\bibnamefont {Kunz}}, \ and\ \bibinfo {author}
  {\bibfnamefont {P.}~\bibnamefont {Pani}},\ }\href {\doibase
  10.1103/PhysRevD.94.104024} {\bibfield  {journal} {\bibinfo  {journal} {Phys.
  Rev. D}\ }\textbf {\bibinfo {volume} {94}},\ \bibinfo {pages} {104024}
  (\bibinfo {year} {2016})},\ \Eprint {http://arxiv.org/abs/1609.01286}
  {arXiv:1609.01286 [gr-qc]} \BibitemShut {NoStop}%
\bibitem [{\citenamefont {Carson}\ and\ \citenamefont
  {Yagi}(2020{\natexlab{a}})}]{Carson:2020cqb}%
  \BibitemOpen
  \bibfield  {author} {\bibinfo {author} {\bibfnamefont {Z.}~\bibnamefont
  {Carson}}\ and\ \bibinfo {author} {\bibfnamefont {K.}~\bibnamefont {Yagi}},\
  }\href {\doibase 10.1088/1361-6382/aba221} {\bibfield  {journal} {\bibinfo
  {journal} {Class. Quant. Grav.}\ }\textbf {\bibinfo {volume} {37}},\ \bibinfo
  {pages} {215007} (\bibinfo {year} {2020}{\natexlab{a}})},\ \Eprint
  {http://arxiv.org/abs/2002.08559} {arXiv:2002.08559 [gr-qc]} \BibitemShut
  {NoStop}%
\bibitem [{\citenamefont {Carson}\ and\ \citenamefont
  {Yagi}(2020{\natexlab{b}})}]{Carson:2020ter}%
  \BibitemOpen
  \bibfield  {author} {\bibinfo {author} {\bibfnamefont {Z.}~\bibnamefont
  {Carson}}\ and\ \bibinfo {author} {\bibfnamefont {K.}~\bibnamefont {Yagi}},\
  }\href {\doibase 10.1103/PhysRevD.101.104030} {\bibfield  {journal} {\bibinfo
   {journal} {Phys. Rev. D}\ }\textbf {\bibinfo {volume} {101}},\ \bibinfo
  {pages} {104030} (\bibinfo {year} {2020}{\natexlab{b}})},\ \Eprint
  {http://arxiv.org/abs/2003.00286} {arXiv:2003.00286 [gr-qc]} \BibitemShut
  {NoStop}%
\bibitem [{\citenamefont {Bonilla}\ \emph {et~al.}(2023)\citenamefont
  {Bonilla}, \citenamefont {Kumar},\ and\ \citenamefont
  {Teukolsky}}]{Bonilla:2022dyt}%
  \BibitemOpen
  \bibfield  {author} {\bibinfo {author} {\bibfnamefont {G.~S.}\ \bibnamefont
  {Bonilla}}, \bibinfo {author} {\bibfnamefont {P.}~\bibnamefont {Kumar}}, \
  and\ \bibinfo {author} {\bibfnamefont {S.~A.}\ \bibnamefont {Teukolsky}},\
  }\href {\doibase 10.1103/PhysRevD.107.024015} {\bibfield  {journal} {\bibinfo
   {journal} {Phys. Rev. D}\ }\textbf {\bibinfo {volume} {107}},\ \bibinfo
  {pages} {024015} (\bibinfo {year} {2023})},\ \Eprint
  {http://arxiv.org/abs/2203.14026} {arXiv:2203.14026 [gr-qc]} \BibitemShut
  {NoStop}%
\bibitem [{\citenamefont {Husa}\ \emph {et~al.}(2016)\citenamefont {Husa},
  \citenamefont {Khan}, \citenamefont {Hannam}, \citenamefont {P\"urrer},
  \citenamefont {Ohme}, \citenamefont {Jim\'enez~Forteza},\ and\ \citenamefont
  {Boh\'e}}]{Husa:2015iqa}%
  \BibitemOpen
  \bibfield  {author} {\bibinfo {author} {\bibfnamefont {S.}~\bibnamefont
  {Husa}}, \bibinfo {author} {\bibfnamefont {S.}~\bibnamefont {Khan}}, \bibinfo
  {author} {\bibfnamefont {M.}~\bibnamefont {Hannam}}, \bibinfo {author}
  {\bibfnamefont {M.}~\bibnamefont {P\"urrer}}, \bibinfo {author}
  {\bibfnamefont {F.}~\bibnamefont {Ohme}}, \bibinfo {author} {\bibfnamefont
  {X.}~\bibnamefont {Jim\'enez~Forteza}}, \ and\ \bibinfo {author}
  {\bibfnamefont {A.}~\bibnamefont {Boh\'e}},\ }\href {\doibase
  10.1103/PhysRevD.93.044006} {\bibfield  {journal} {\bibinfo  {journal} {Phys.
  Rev. D}\ }\textbf {\bibinfo {volume} {93}},\ \bibinfo {pages} {044006}
  (\bibinfo {year} {2016})},\ \Eprint {http://arxiv.org/abs/1508.07250}
  {arXiv:1508.07250 [gr-qc]} \BibitemShut {NoStop}%
\bibitem [{\citenamefont {Khan}\ \emph {et~al.}(2016)\citenamefont {Khan},
  \citenamefont {Husa}, \citenamefont {Hannam}, \citenamefont {Ohme},
  \citenamefont {P\"urrer}, \citenamefont {Jim\'enez~Forteza},\ and\
  \citenamefont {Boh\'e}}]{Khan:2015jqa}%
  \BibitemOpen
  \bibfield  {author} {\bibinfo {author} {\bibfnamefont {S.}~\bibnamefont
  {Khan}}, \bibinfo {author} {\bibfnamefont {S.}~\bibnamefont {Husa}}, \bibinfo
  {author} {\bibfnamefont {M.}~\bibnamefont {Hannam}}, \bibinfo {author}
  {\bibfnamefont {F.}~\bibnamefont {Ohme}}, \bibinfo {author} {\bibfnamefont
  {M.}~\bibnamefont {P\"urrer}}, \bibinfo {author} {\bibfnamefont
  {X.}~\bibnamefont {Jim\'enez~Forteza}}, \ and\ \bibinfo {author}
  {\bibfnamefont {A.}~\bibnamefont {Boh\'e}},\ }\href {\doibase
  10.1103/PhysRevD.93.044007} {\bibfield  {journal} {\bibinfo  {journal} {Phys.
  Rev. D}\ }\textbf {\bibinfo {volume} {93}},\ \bibinfo {pages} {044007}
  (\bibinfo {year} {2016})},\ \Eprint {http://arxiv.org/abs/1508.07253}
  {arXiv:1508.07253 [gr-qc]} \BibitemShut {NoStop}%
\bibitem [{\citenamefont {Garc\'\i{}a-Quir\'os}\ \emph
  {et~al.}(2020)\citenamefont {Garc\'\i{}a-Quir\'os}, \citenamefont {Colleoni},
  \citenamefont {Husa}, \citenamefont {Estell\'es}, \citenamefont {Pratten},
  \citenamefont {Ramos-Buades}, \citenamefont {Mateu-Lucena},\ and\
  \citenamefont {Jaume}}]{Garcia-Quiros:2020qpx}%
  \BibitemOpen
  \bibfield  {author} {\bibinfo {author} {\bibfnamefont {C.}~\bibnamefont
  {Garc\'\i{}a-Quir\'os}}, \bibinfo {author} {\bibfnamefont {M.}~\bibnamefont
  {Colleoni}}, \bibinfo {author} {\bibfnamefont {S.}~\bibnamefont {Husa}},
  \bibinfo {author} {\bibfnamefont {H.}~\bibnamefont {Estell\'es}}, \bibinfo
  {author} {\bibfnamefont {G.}~\bibnamefont {Pratten}}, \bibinfo {author}
  {\bibfnamefont {A.}~\bibnamefont {Ramos-Buades}}, \bibinfo {author}
  {\bibfnamefont {M.}~\bibnamefont {Mateu-Lucena}}, \ and\ \bibinfo {author}
  {\bibfnamefont {R.}~\bibnamefont {Jaume}},\ }\href {\doibase
  10.1103/PhysRevD.102.064002} {\bibfield  {journal} {\bibinfo  {journal}
  {Phys. Rev. D}\ }\textbf {\bibinfo {volume} {102}},\ \bibinfo {pages}
  {064002} (\bibinfo {year} {2020})},\ \Eprint
  {http://arxiv.org/abs/2001.10914} {arXiv:2001.10914 [gr-qc]} \BibitemShut
  {NoStop}%
\bibitem [{\citenamefont {Blanchet}(2014)}]{Blanchet:2013haa}%
  \BibitemOpen
  \bibfield  {author} {\bibinfo {author} {\bibfnamefont {L.}~\bibnamefont
  {Blanchet}},\ }\href {\doibase 10.12942/lrr-2014-2} {\bibfield  {journal}
  {\bibinfo  {journal} {Living Rev. Rel.}\ }\textbf {\bibinfo {volume} {17}},\
  \bibinfo {pages} {2} (\bibinfo {year} {2014})},\ \Eprint
  {http://arxiv.org/abs/1310.1528} {arXiv:1310.1528 [gr-qc]} \BibitemShut
  {NoStop}%
\bibitem [{\citenamefont {Yunes}\ \emph {et~al.}(2010)\citenamefont {Yunes},
  \citenamefont {Pretorius},\ and\ \citenamefont {Spergel}}]{Yunes:2009bv}%
  \BibitemOpen
  \bibfield  {author} {\bibinfo {author} {\bibfnamefont {N.}~\bibnamefont
  {Yunes}}, \bibinfo {author} {\bibfnamefont {F.}~\bibnamefont {Pretorius}}, \
  and\ \bibinfo {author} {\bibfnamefont {D.}~\bibnamefont {Spergel}},\ }\href
  {\doibase 10.1103/PhysRevD.81.064018} {\bibfield  {journal} {\bibinfo
  {journal} {Phys. Rev. D}\ }\textbf {\bibinfo {volume} {81}},\ \bibinfo
  {pages} {064018} (\bibinfo {year} {2010})},\ \Eprint
  {http://arxiv.org/abs/0912.2724} {arXiv:0912.2724 [gr-qc]} \BibitemShut
  {NoStop}%
\bibitem [{\citenamefont {Cabero}\ \emph {et~al.}(2017)\citenamefont {Cabero},
  \citenamefont {Nielsen}, \citenamefont {Lundgren},\ and\ \citenamefont
  {Capano}}]{Cabero:2016ayq}%
  \BibitemOpen
  \bibfield  {author} {\bibinfo {author} {\bibfnamefont {M.}~\bibnamefont
  {Cabero}}, \bibinfo {author} {\bibfnamefont {A.~B.}\ \bibnamefont {Nielsen}},
  \bibinfo {author} {\bibfnamefont {A.~P.}\ \bibnamefont {Lundgren}}, \ and\
  \bibinfo {author} {\bibfnamefont {C.~D.}\ \bibnamefont {Capano}},\ }\href
  {\doibase 10.1103/PhysRevD.95.064016} {\bibfield  {journal} {\bibinfo
  {journal} {Phys. Rev. D}\ }\textbf {\bibinfo {volume} {95}},\ \bibinfo
  {pages} {064016} (\bibinfo {year} {2017})},\ \Eprint
  {http://arxiv.org/abs/1602.03134} {arXiv:1602.03134 [gr-qc]} \BibitemShut
  {NoStop}%
\bibitem [{\citenamefont {Finn}(1992)}]{Finn:1992wt}%
  \BibitemOpen
  \bibfield  {author} {\bibinfo {author} {\bibfnamefont {L.~S.}\ \bibnamefont
  {Finn}},\ }\href {\doibase 10.1103/PhysRevD.46.5236} {\bibfield  {journal}
  {\bibinfo  {journal} {Phys. Rev. D}\ }\textbf {\bibinfo {volume} {46}},\
  \bibinfo {pages} {5236} (\bibinfo {year} {1992})},\ \Eprint
  {http://arxiv.org/abs/gr-qc/9209010} {arXiv:gr-qc/9209010} \BibitemShut
  {NoStop}%
\bibitem [{\citenamefont {Biwer}\ \emph {et~al.}(2019)\citenamefont {Biwer},
  \citenamefont {Capano}, \citenamefont {De}, \citenamefont {Cabero},
  \citenamefont {Brown}, \citenamefont {Nitz},\ and\ \citenamefont
  {Raymond}}]{Biwer:2018osg}%
  \BibitemOpen
  \bibfield  {author} {\bibinfo {author} {\bibfnamefont {C.~M.}\ \bibnamefont
  {Biwer}}, \bibinfo {author} {\bibfnamefont {C.~D.}\ \bibnamefont {Capano}},
  \bibinfo {author} {\bibfnamefont {S.}~\bibnamefont {De}}, \bibinfo {author}
  {\bibfnamefont {M.}~\bibnamefont {Cabero}}, \bibinfo {author} {\bibfnamefont
  {D.~A.}\ \bibnamefont {Brown}}, \bibinfo {author} {\bibfnamefont {A.~H.}\
  \bibnamefont {Nitz}}, \ and\ \bibinfo {author} {\bibfnamefont
  {V.}~\bibnamefont {Raymond}},\ }\href {\doibase 10.1088/1538-3873/aaef0b}
  {\bibfield  {journal} {\bibinfo  {journal} {Publ. Astron. Soc. Pac.}\
  }\textbf {\bibinfo {volume} {131}},\ \bibinfo {pages} {024503} (\bibinfo
  {year} {2019})},\ \Eprint {http://arxiv.org/abs/1807.10312} {arXiv:1807.10312
  [astro-ph.IM]} \BibitemShut {NoStop}%
\bibitem [{\citenamefont {Foreman-Mackey}\ \emph {et~al.}(2013)\citenamefont
  {Foreman-Mackey}, \citenamefont {Hogg}, \citenamefont {Lang},\ and\
  \citenamefont {Goodman}}]{Foreman_Mackey_2013}%
  \BibitemOpen
  \bibfield  {author} {\bibinfo {author} {\bibfnamefont {D.}~\bibnamefont
  {Foreman-Mackey}}, \bibinfo {author} {\bibfnamefont {D.~W.}\ \bibnamefont
  {Hogg}}, \bibinfo {author} {\bibfnamefont {D.}~\bibnamefont {Lang}}, \ and\
  \bibinfo {author} {\bibfnamefont {J.}~\bibnamefont {Goodman}},\ }\href
  {\doibase 10.1086/670067} {\bibfield  {journal} {\bibinfo  {journal}
  {Publications of the Astronomical Society of the Pacific}\ }\textbf {\bibinfo
  {volume} {125}},\ \bibinfo {pages} {306} (\bibinfo {year}
  {2013})}\BibitemShut {NoStop}%
\bibitem [{\citenamefont {Collaboration}\ \emph {et~al.}(2021)\citenamefont
  {Collaboration}, \citenamefont {the Virgo~Collaboration},\ and\ \citenamefont
  {the KAGRA~Collaboration}}]{collaboration2021gwtc3}%
  \BibitemOpen
  \bibfield  {author} {\bibinfo {author} {\bibfnamefont {T.~L.~S.}\
  \bibnamefont {Collaboration}}, \bibinfo {author} {\bibnamefont {the
  Virgo~Collaboration}}, \ and\ \bibinfo {author} {\bibnamefont {the
  KAGRA~Collaboration}},\ }\href@noop {} {\  (\bibinfo {year} {2021})},\
  \Eprint {http://arxiv.org/abs/2111.03606} {arXiv:2111.03606 [gr-qc]}
  \BibitemShut {NoStop}%
\end{thebibliography}%

\end{document}